\documentclass[reprint,showpacs,prd,preprintnumbers,amssymb,nofootinbib,superscriptaddress,aps]{revtex4-2}
\pdfoutput=1
\usepackage{amsmath,amssymb,bm}
\usepackage{bbold}
\usepackage{lipsum}
\usepackage{hyperref}
\usepackage{graphics,graphicx}
\usepackage{subcaption}
\usepackage{csquotes}
\usepackage{mathtools}
\usepackage{slashed}
\usepackage{color,colordvi}
\usepackage{soul}
\usepackage{xcolor}
\usepackage{array}
\usepackage{multirow}
\usepackage{silence}
\usepackage{booktabs}
\usepackage{multirow}
\usepackage{tabularx,ragged2e}
\usepackage{comment}
\usepackage{slashed}
\usepackage{float}
\usepackage[normalem]{ulem}
\renewcommand{\(}{\left(}
\renewcommand{\)}{\right)}
\renewcommand{\{}{\left\lbrace}
\renewcommand{\}}{\right\rbrace}

\newcommand{\nn}{\nonumber}

\newcommand{\order}[1]{\mathcal{O}\({#1}\)}
\newcommand{\GeV}{\,\mathrm{GeV}}

\newcommand{\MeV}{\,\mathrm{MeV}}

\newcommand{\Mt}{M^2_{\tau}}
\newcommand{\hs}{\hspace{.4mm}}
\newcommand{\bs}{\hspace{1cm}}

\newcommand{\mtsq}{M_{\tau}^2}
\begin{document}
	\title{Renormalization group improved \texorpdfstring{$m_s$}{} and \texorpdfstring{$\vert V_{us}\vert$}{} determination from hadronic \texorpdfstring{$\tau$ }{}decays.}
	\author{B. Ananthanarayan}
	\email{anant@iisc.ac.in}
	\affiliation{Centre for High Energy Physics, Indian Institute of Science, Bangalore 560 012, India}
	\author{Diganta Das}
	\email{diganta.das@iiit.ac.in}
	\affiliation{Center for Computational Natural Sciences and Bioinformatics, International Institute of Information
		Technology, Hyderabad 500 032, India}
	\author{M. S. A. Alam Khan}
	\email{alam.khan1909@gmail.com}
	\affiliation{Centre for High Energy Physics, Indian Institute of Science, Bangalore 560 012, India}
	\begin{abstract}
		We determine the strange quark mass (\texorpdfstring{$m_s$}{}) and quark mixing element \texorpdfstring{$\vert V_{us}\vert $}{}, and their joint determination from the Cabibbo suppressed hadronic $\tau$ decays in various perturbative schemes. We have improved this analysis compared to the previous analysis based on the optimal renormalization or the renormalization group summed perturbation theory (RGSPT) scheme, by replacing the theoretical longitudinal contributions with phenomenological parametrization, and the RGSPT coefficients are used for the dimension-4 Adler functions. The improved analysis results in the extraction of \texorpdfstring{$m_s(2\GeV)=98\pm19\MeV$}{} and \texorpdfstring{$\vert V_{us}\vert=0.2191\pm0.0043$}{} from the RGSPT scheme.
	\end{abstract}
	\maketitle
	\section{Introduction}
	The hadronic decays of the $\tau$ leptons have been of constant interest for determining various parameters of the Standard Model (SM) of particle physics. The availability of experimental data on the strange and non-strange decay modes for the hadronic $\tau$ decays has opened the window for the determination of various parameters relevant for the quantum chromodynamics (QCD), namely the strong coupling constant $\alpha_s$, the strange quark mass $m_s$, the vacuum condensates, the low-energy chiral couplings, and the quark mixing element $\vert V_{us}\vert$ of the Cabibbo–Kobayashi–Maskawa (CKM) matrix (see ref.~\cite{Davier:2005xq,Pich:2013lsa} for details). \par
	On the theoretical side, the QCD contributions to the hadronic $\tau$ decays are studied by evaluating the current correlator using the Operator Product Expansion (OPE)\cite{Wilson:1969zs}. The OPE factorizes the long- and short-distance contributions. The long-distance information is encoded into the vacuum condensates. The short distance part is written as the perturbative series in the strong coupling constant and quark masses. The vacuum condensates can also be evaluated using chiral perturbation theory (ChPT) \cite{Gasser:1984gg}, lattice QCD \cite{FlavourLatticeAveragingGroup:2019iem}, and Renormalization Group (RG) optimized perturbation theory \cite{Kneur:2015dda,Kneur:2020bph}. The short distance contributions require the evaluation of the Feynman diagrams. It is also known that some of the contributions to the hadronic vacuum polarization function are not captured by the OPE, and it is a quark-hadron duality violation. These duality violating terms are parameterized in a model-dependent way and fitted to experimental data and should also be added to the OPE contributions \cite{Boito:2017cnp}.\par
	The longitudinal component of the QCD Adler function, corresponding to the zero angular momentum state, has been calculated to $\order{\alpha_s^4}$\cite{Becchi:1980vz,Broadhurst:1981jk,Chetyrkin:1996sr,Baikov:2005rw,Gorishnii:1990zu,Gorishnii:1991zr}. It has poor convergence behavior and raises the question of the method's applicability in the extraction of strange quark mass. This problem can be cured by replacing these contributions with their phenomenological input and has been used in ref.~\cite{Gamiz:2002nu,Jamin:2001zq,Jamin:2001zr,Maltman:2001gc,Maltman:2001sv} for $m_s$ and $\vert V_{us}\vert$ extractions from the experimental moment data. These improvements have resulted in much better control over the theoretical uncertainties in the $m_s$ and $\vert V_{us}\vert$ determinations.\par
	The hadronic $\tau$ decays have been extensively studied using various perturbative schemes. These schemes differ in how strong coupling constant and quark masses are evaluated along the contour in the complex plane using their renormalization-group (RG) properties. The most commonly used schemes in the extraction of strange quark mass and CKM matrix element from the Cabibbo suppressed hadronic $\tau$ decay are fixed-order perturbation theory (FOPT) and contour-improved perturbation theory (CIPT). For the hadronic $\tau$ decays, the FOPT suffers from the problem of large logarithms along the contour in the complex energy plane, and the higher-order spectral moments are very sensitive to scale variations. In the CIPT scheme, direct numerical evaluation of coupling constant and masses along complex contour using their RGE does not suffer from the problem of the large logarithms. However, scale dependence is still the major source of theoretical uncertainties for higher moments.\par Recently, the optimal renormalization or RG-summed perturbation theory (RGSPT) has been used by two of us in ref.~\cite{Ananthanarayan:2016kll} in the strange quark mass determination. The behavior of polarization and Alder functions in the complex contour was also studied for RGSPT, CIPT, FOPT, and the method of effective charges (MEC) in great detail. However, the numerical impact of the theoretical uncertainties from perturbation series truncation, and scale dependence were excluded. We improve the previous analysis by:
	\begin{itemize}
		\item Including the RGSPT coefficients for dimension-4 Adler functions.
		\item Replacing the divergent longitudinal perturbative QCD expressions for the Adler function with the phenomenological parametrization used in ref.~\cite{Gamiz:2002nu,Maltman:2001gc}. This replacement significantly reduces the theoretical uncertainties.
		\item Performing $\vert V_{us}\vert$ as well as the joint $m_s$ and $\vert V_{us}\vert$ determinations for the first time using RGSPT.
		\item The effects of the variation of the $m_s$ and $|V_{us}|$ with the variation of the moments calculated at different energies ($s_0<\mtsq$) is also included and found to be constituting an important source of uncertainty. 
		\item Using the five-loop QCD $\beta-$function and anomalous dimensions for the running of the strong coupling constant and quark masses.
	\end{itemize}
	\par
	The article is organized as follows: Section~\ref{sec:formalism} provides a brief overview of the various quantities that are needed for the extraction of $m_s$ and $\vert V_{us}\vert$. A short introduction to RGSPT is given in section~\ref{sec:RGSPT}. Section~\ref{sec:OPE_contributions} explains the OPE contributions to the Adler function. The behavior of leading-order mass corrections to the Adler functions in different schemes used in this article is studied in the section~\ref{sec:dim_2_behaviour}. The higher-order term of the perturbation series becomes very important for the higher moments, and two prescriptions for the truncation of the perturbation series are also defined in this section. In section~\ref{sec:rev_pheno}, the phenomenological parametrization of longitudinal contributions is briefly discussed. Then, we move to section~\ref{sec:ms_pert} where strange quark mass is extracted using only the perturbative QCD (pQCD) contributions calculated from OPE. The weighted average results for the $m_s(\mtsq)$ extraction  using this method in CIPT, FOPT and RGSPT schemes are presented in table~\ref{tab:mspertweighted}. The details of uncertainties can be found in the appendix~\ref{app:pQCD_mass}. In section ~\ref{sec:pheno_ms} the $m_s(\mtsq)$ determination using the phenomenological parametrization for the longitudinal component is performed and results are presented in table~\ref{tab:msphenoweighted2}. Details of the strange quark mass determinations from moments are presented in the appendix~\ref{app:pheno_mass}. The determination of $\vert V_{us}\vert$ using external input for $m_s$ is performed in section~\ref{sec:Vusextraction}. The weighted average results using the OPAL  and ALEPH data are presented in table~\ref{tab:Vus_weightedmean12} and table~\ref{tab:Vus_Aleph1}, respectively. The details of determinations from the moments as well as the uncertainties coming from various sources are presented in appendix~\ref{app:vus}. In section~\ref{sec:joint_msVus} the joint extraction of $m_s$ and $\vert V_{us}\vert$ is performed. We provide a summary and conclusion in section~\ref{sec:summary}. We also provide supplementary inputs needed for this article in the appendix~\ref{app:mass_run},\ref{app:summed_sol},\ref{app:adlercoef},\ref{app:dim4corrections}. Details of $m_s$ and $\vert V_{us}\vert$ determinations from moments can be found in appendix~\ref{app:determination_details}.
	\section{Formalism}\label{sec:formalism}
	An important quantity for the study of hadronic $\tau$ decay width\cite{Braaten:1991qm,Pich:1999hc,Narison:1988ni} is the two-point current correlator:
	\begin{equation}
		\Pi^{V/A}_{\mu \nu,ij} (p^2) \equiv i \int dy \text{ }e^{i p y}\text{ }\langle\Omega\left|T\{J^{V/A}_{\mu,ij}\left(y\right)J^{V/A}_{\nu,ij}\left(0\right)^{\dagger}\}\right|\Omega\rangle
		\label{cor}
	\end{equation}
	where $|\Omega\rangle$ denotes the physical vacuum, $J^{V/A}_{\mu,ij}\left(y\right)=\left(\bar{q}_j\gamma_{\mu}/\left(\gamma_{\mu}\gamma_{5}\right)q_i\right)\left(x\right)$ is the hadronic vector/axial current, and the indices $i$ and $j$ denote the flavors of light quarks. The current correlator can be calculated perturbatively using OPE \cite{Wilson:1969zs} as a power expansion in $1/p$, and the corresponding coefficients of are the operator of that dimension. Purely perturbative corrections appear up to dimension-2 in the OPE expansion, and the long-distance corrections corresponding to the vacuum condensates start from dimension-4.\par
	Using Lorentz decomposition, the current correlator in eq.~\eqref{cor} can be decomposed into the longitudinal and transverse components with angular momentum $J=0$ and $J=1$ as:
	\begin{equation}
		\Pi^{V/A}_{\mu \nu,ij} (p^2) =\left(p_{\mu}p_{\nu}-g_{\mu,\nu}\right) \Pi^{V/A,T}_{ij} (p^2)+ p_{\mu} p_{\nu} \Pi^{V/A,L}_{ij}(p^2)\,.
		\label{eq:cor_decompose}
	\end{equation}
	The $L/T$ correlators are related to experimentally measurable semi-hadronic $\tau$ decay rate ($R_{\tau}$), defined by:
	\begin{equation}
		R_{\tau}\equiv\frac{\Gamma\left(\tau^-\rightarrow\text{hadrons } \nu_{\tau}\left(\gamma\right)\right)}{\Gamma\left(\tau^-\rightarrow e^- \nu_{\tau}\left(\gamma\right)\right)}=R_{\tau,V}+R_{\tau,A}+R_{\tau,S}\,,
	\end{equation}
	and it is related to the imaginary part of the current correlators in eq.~\eqref{eq:cor_decompose} by:
	\begin{align}
		R_{\tau}\left(s_0\right)=12 \pi &\int_{0}^{s_0} \frac{ds}{s_0}\left(1-\frac{s}{s_0}\right)^2 \nonumber\\&\times \left[\left(1+\frac{2 s}{s_0}\right) \text{Im}\left(\Pi^T\right)(s)+ \text{Im}\left(\Pi^L\right)(s)\right]\,.
		\label{eq:Rratio}
	\end{align}
	It should be noted that these current correlators also carry the information about mixing among the quark flavors and can be written as:
	\begin{equation}
		\Pi^J\equiv \sum_{i=d,s}\left|V_{ui}\right|^2 \left[\Pi^{V,J}_{ui}\left(s\right)+\Pi^{A,J}_{ui}\left(s\right)\right]\,,\label{R_tau1}
	\end{equation}
	and $\vert V_{ij}\vert$ are the elements of the CKM matrix. \par
	To study the invariant mass distribution of final-state hadrons, we need moments from the hadronic $\tau$ decay rate, defined by\cite{LeDiberder:1992zhd}:
	\begin{align}
		R^{kl}_{\tau}\left(s_0\right)\equiv \int_{0}^{s_0} ds \left(1-\frac{s}{s_0}\right)^k \left(\frac{s}{s_0}\right)^l \frac{d R_{\tau}}{ds}\,,
		\label{eq:rklmomentdef}
	\end{align}
	Using integration by parts, we can convert eq.~\eqref{eq:rklmomentdef} into the following form:
	\begin{align}
		R^{kl}_{\tau}\left(s_0\right)=- i \pi  &\oint\limits_{\left|x_c\right|=1} \frac{dx_c}{x_c}\times\big\lbrace 3 \mathcal{F}^{L+T}_{kl}(x_c) \mathcal{D}^{L+T}\left(s_0 x_c\right)\nonumber\\ &\bs\bs+4 \mathcal{F}^{L}_{kl}(x_c)\mathcal{D}^{L}\left(s_0 x_c\right)\big\rbrace\,,
		\label{Rtau_def}
	\end{align}
	where $x_c=s/s_0$ and $\mathcal{D}^{L+T}(s)$ are the Adler functions. \par Usually, the experimental value of the moments defined above are provided for  $s_0=\mtsq$ in the literature. However, their values at different energy can be calculated using the experimental data on the spectral functions provided in the refs.~\cite{OPAL:1998rrm,OPAL:2004icu,Davier:2013sfa,Boito:2020xli}. \par The Adler function satisfies the homogeneous renormalization group equation (RGE) and is related to the current correlators by the relation:
	\begin{align}
		\mathcal{D}^{L+T}\left(s\right)&\equiv -s\frac{d}{ds}\left(\Pi^{L+T}\left(s\right)\right)\,,\\ \mathcal{D}^{L}\left(s\right)&\equiv \frac{s}{\Mt}\frac{d}{ds}\left(s\hspace{1 mm} \Pi^{L}\left(s\right)\right).
		\label{eq:Ds}
	\end{align}
	The resulting quantity in eq.~\eqref{Rtau_def} is an expansion in the strong coupling constant, quark masses, and condensates of higher dimension operators. It explicitly depends on the CKM matrix element and the electroweak corrections. These terms are not shown in eq.~\eqref{Rtau_def} but factored out in eq.~\eqref{eq:Rtauexpanded} and eq.~\eqref{eq:delRkl}.
	The kinematic kernels $ \mathcal{F}^{kl}_{L+T}\left(x_c\right)$ and $ \mathcal{F}^{kl}_{L}\left(x_c\right)$ appearing in the eq.~\eqref{Rtau_def} are given by:
	\begin{align}
		\mathcal{F}^{kl}_{L+T}(x_c)\equiv &2\left(1-x_c\right)^{3+k} \sum_{n=0}^{l} \frac{l!}{\left(l-n\right)! n!}\left(x_c-1\right)^n\times \nonumber\\&\bs\hs\hs\frac{(6+k+n)+2(3+k+n) x_c}{(3+k+n)(4+k+n)}\,,\\
		\mathcal{F}^{kl}_{L}(x_c)\equiv &3\left(1-x_c\right)^{3+k} \sum_{n=0}^{l} \frac{l!}{\left(l-n\right)! n!}\frac{\left(x_c-1\right)^n} {(3+k+n)}\,,
	\end{align}
	and their explicit form used in this article is presented in the table~\ref{tab:klmoments}.
	\begin{table}
		\begin{center}
			\begin{tabular}{|c|c|c|}	\hline
				$(k, l)$ & $\mathcal{F}_{L+T}^{k l}(x)$ & $\mathcal{F}_{L}^{k l}(x)$ \\	\hline
				(0,0) & $(1-x)^{3}(1+x)$ & $(1-x)^{3}$ \\
				(1,0) & $\frac{1}{10}(1-x)^{4}(7+8 x)$ & $\frac{3}{4}(1-x)^{4}$ \\
				(2,0) & $\frac{2}{15}(1-x)^{5}(4+5 x)$ & $\frac{3}{5}(1-x)^{5}$ \\
				(3,0) & $\frac{1}{7}(1-x)^{6}(3+4 x)$ & $\frac{1}{2}(1-x)^{6}$ \\
				(4,0) & $\frac{1}{14}(1-x)^{7}(5+7 x)$ & $\frac{3}{7}(1-x)^{7}$\\ \hline
			\end{tabular}
			\caption{Kinematic kernels used in this article.}
			\label{tab:klmoments}
		\end{center}
	\end{table}
	\par
	Performing the contour integral defined in eq.~\eqref{Rtau_def}, we can write $R^{kl}_{\tau}$\cite{Braaten:1991qm} as:
	\begin{align}  \label{eq:Rtauexpanded}
		R^{kl}_{\tau}\left(s_0\right)&=3 \left(\vert V_{ud}\vert^2+\vert V_{us}\vert ^2\right) S_{EW} \bigg\lbrace1+\delta'_{EW}+\delta^{\left(0\right),kl}\nonumber\\&+\sum_{n=2,4...}\left(\cos^2\left(\theta_C\right) \delta^{\left(n\right),kl}_{ud}+ \sin^2 \left(\theta_C\right) \delta^{\left(n\right),kl}_{us}\right)\bigg\rbrace\,,
	\end{align}
	where $\theta_C=\sin^{-1}\left(\vert V_{us}\vert/\sqrt{\left(|V_{us}|^2+|V_{ud}|^2\right)}\right)$ is the Cabibbo angle, $\delta^{(n)}_{ud,us}$\cite{Braaten:1991qm} carry the information of contour integrals evaluated in eq.~\eqref{Rtau_def},  $\delta'_{EW}=0.0010$ and $S_{EW}=1.0201\pm0.0003$ are one-loop RG-improved electroweak corrections \cite{Braaten:1990ef,Erler:2002mv}.\par    The most important quantity of interest in the determination of the strange quark mass is the $SU(3)$ breaking terms $\delta R^{kl}_{\tau}\left(s_0\right)$\cite{ALEPH:1999uux} defined as
	\begin{align}  \label{eq:delRkl}
		\delta R^{kl}_{\tau}\left(s_0\right)&\equiv\frac{R^{kl}_{\tau,V+A}\left(s_0\right)}{\left\vert V_{ud}\right\vert^2} -\frac{R^{kl}_{\tau,S}\left(s_0\right)}{\left\vert V_{us}\right\vert^2} \\&=3 S_{EW}\sum_{n\ge2}\left(\delta^{\left(n\right),kl}_{ud}-\delta^{\left(n\right),kl}_{us}\right)\,,
	\end{align}
	which is free from instanton and renormalon contributions and vanishes in the chiral limit. It is an experimentally measurable quantity that is used as input from table~\ref{tab:rkl_exp} along with the theoretical quantities appearing in eq.~\eqref{eq:delRkl} in the strange quark mass determination. \par
	The value of strong coupling constant $\alpha_s(M^2_Z)=0.1179\pm.0010$ has been taken from ref.~\cite{Zyla:2020zbs} and evolved to $\tau$ lepton mass scale using five-loop $\beta-$function using package REvolver \cite{Hoang:2021fhn}.  Its value at $\tau$ lepton mass is $\alpha_s(M^2_{\tau})=0.3187\pm 0.0083$ and has been used in this article.
	\section{Review of Optimal renormalization}\label{sec:RGSPT}
	The optimal renormalization technique is used to resum the running logarithms present in the perturbation series using RGE\cite{Ahmady:1999xg,Ahmady:2002fd,Ahmady:2002pa}. The resulting summed series shows reduced scale dependence and hence a reduction in the theoretical uncertainty in the extraction of a quantity of interest. In the case of hadronic $\tau$ decays, where weighted integrals along the complex contour are involved, these running logarithms become very important. Their summation is necessary to perform the perturbative analysis properly. RGSPT resum these logarithms, and the resulting fixed order truncated series has less sensitivity to scale variations even for higher moments than FOPT and CIPT.\par
	The perturbative series describing a QCD process is given by:
	\begin{align}
		W(x,m)&=x^{n_1} m^{n_2} \sum_{i=0}x^i L^j \hspace{.4mm}T_{i,j}\,,	\label{eq:series_expanded}
	\end{align}
	where $x\equiv x(\mu^2)$, $L\equiv \log(\mu^2/q^2)$ and $m=m(\mu^2)$. We can rewrite the series as follows:
	\begin{align}
		W^{RG\Sigma}=x^{n_1} m^{n_2}\sum_{i=0}x^i\hspace{.4mm}S_{i}[x \hspace{.4 mm} L]\,,
		\label{ser_summed}
	\end{align}
	where the $S_{i}[x L]$-coefficients are given by:
	\begin{align}
		S_{i}[x\hspace{.4mm}L]=\sum_{n=i}^{\infty} T_{n,n-i}  (x\hspace{.4mm}L)^{n-i}\,.
		\label{eq:summed_coefs}
	\end{align}
	The RGE for eq.~\eqref{eq:series_expanded} is given by:
	\begin{align}
		&\mu^2 \frac{d}{d\mu^2} W(x,m)-\gamma_{a}(x) W(x,m)=0\,,\\&\left(\beta(x) \partial_x+ \gamma_m(x) \partial_m+\partial_L-\gamma_{a}(x)\right)W(x,m)=0\,,
	\end{align}
	where $\gamma_{a}(x)=\sum_{i=0}x^{i+1}\gamma_{a}^{\left(i\right)}$ is the anomalous dimension associated with  $W(x,m)$. We can collect the terms corresponding to summed coefficients defined in eq.~\eqref{eq:summed_coefs}. This process results in a set of coupled differential equations for $S_{i}[w]$, which can be summarized as:
	\begin{align}
		\sum _{i=0}^n \bigg[\beta _i &(\delta_{i,0}+w-1)  S_{n-i}'(w)\nonumber\\&+S_{n-i}(w) \left(n_2 \gamma_i+\beta _i (n_1-i+n)+\gamma_a^{(i)}\right)\bigg]=0\,.
		\label{summed_de}
	\end{align}
	Here we have substituted $w=1-\beta_0 x L$, which simplifies the solutions of differential equations. For further details on RG summation, we refer to ref.~\cite{Ahmady:1999xg,Ahmady:2002fd,Ahmady:2002pa,Abbas:2012py,Ananthanarayan:2016kll,Ananthanarayan:2020umo,Abbas:2022wnz}. \par
	The solution to the first three summed coefficients, relevant for dimension-0 and dimension-2 Adler functions, appearing in the eq.~\eqref{summed_de} can be found in the appendix~\ref{app:summed_sol}. It should be noted that the RGE for dimension-4 operators mix perturbative coefficients with condensates; hence they do not obey eq.~\eqref{ser_summed}. The RG-summed perturbative coefficients are relevant for eq.~\eqref{eq:dim4LTAdler} and eq.~\eqref{eq:dim4LAdler} in the appendix~\ref{app:dim4corrections}.
	
	\section{OPE Contributions to the QCD Adler Function}\label{sec:OPE_contributions}
	\subsection{Leading order contribution}
	Dimension-zero is the leading perturbative contribution to the current correlator in the massless limit and has been calculated to $\order{\alpha_s^4}$ \cite{Appelquist:1973uz,Zee:1973sr,Chetyrkin:1979bj,Dine:1979qh,Gorishnii:1990vf,Surguladze:1990tg,Chetyrkin:1996ez,Baikov:2008jh,Baikov:2010je,Herzog:2017dtz}. It receives a contribution only from the transverse piece of the current correlator, which is identical for both vector and axial-vector channels and thus cancels in the eq.~\eqref{eq:delRkl}. The Adler functions obtained using the OPE can be organized as follows:
	\begin{align}
		\mathcal{D}^{L+T}(s)&=\sum_{n=0,2,4,\cdots}\frac{1}{s^\frac{n}{2}}\mathcal{D}_{n}^{L+T}(s)\,,\\
		\mathcal{D}^{L}(s)&=\frac{1}{M_{\tau}^2}\sum_{n=2,4,\cdots} \frac{1}{s^{\frac{n}{2}-1}}\mathcal{D}_{n}^{L}(s)\,,
	\end{align}
	where $ij$ are the flavor indices and the Adler functions in the RHS of the above equations are expansion in $\alpha_s$, $m_q$, and the quark and gluon condensate terms. Their definition gets clearer if we take a contour integration along $s=\mtsq e^{i \phi}$ and the coefficient of $(\mtsq)^{-n}$ are called operators of dimension $2n$.
	The massless Adler functions are given by
	\begin{align}
		\mathcal{D}_{0}^{L+T,V/A}(s)&=\frac{1}{4\pi}\sum_{i}x(-s)^i \tilde{K}^{L+T}_{i}\,,\\
		\mathcal{D}_{0}^{L,V/A}(s)&=0\,,
	\end{align}
	where $x(-s)=x(q^2)=\alpha_s(q^2)/\pi$ and $\tilde{K}^{L+T}_{i}$ are the coefficient of Adler function at $i^{th}-$loop which can be found in appendix~\ref{app:adlercoef}. The RG running of dimension-zero $``L+T"-$component of the Adler function is given by:
	\begin{align}
		\mu^2 \frac{d}{d\hspace{.4 mm} \mu^2}&\mathcal{D}^{ L+T,V/A}_0(s)\nonumber\\&=\Bigg(\beta(x) \frac{\partial}{\partial x}+\frac{\partial}{\partial L}\Bigg) \mathcal{D}^{L+T,V/A}_0(s)=0\,,
	\end{align}
	where $L=\log(\frac{\mu^2}{-s})$, and the QCD beta function ($\beta(x)$) is defined as:
	\begin{align}
		\mu^2\frac{d}{d\mu^2}x(\mu^2)=\beta(x(\mu^2))=-\sum_i \beta_i x(\mu^2)^{i+2} \,. \label{beta_function}
	\end{align}
	The coefficients of the beta function $\beta_i$'s are known up to the five-loops and are presented in the appendix~\ref{app:mass_run}.
	
	\subsection{The Dimension-2 contributions to the Adler Function}
	The leading  order mass corrections to hadronic $\tau$ decay rate come from the dimension-2 Adler function. The $\mathcal{D}^{L+T}(s)$ Adler function is known to be $\order{\alpha_s^3}$. \cite{Baikov:2004ku,Baikov:2004tk,Chetyrkin:1993hi,Gorishnii:1986pz,Generalis:1989hf,Bernreuther:1981sp} while $\mathcal{D}^{L}$ is known to $\order{\alpha_s^4}$ \cite{Becchi:1980vz,Broadhurst:1981jk,Chetyrkin:1996sr,Baikov:2005rw,Gorishnii:1990zu,Gorishnii:1991zr} and their analytic expression can be found in the appendix~\ref{app:dim2adler}.
	The RG running of dimension-2 operators is given by:
	\begin{align}
		\mu^2 \frac{d}{d\hspace{.4 mm} \mu^2}\mathcal{D}^{J}_2(s)=&\Bigg\lbrace \frac{\partial}{\partial L}+\beta(x(\mu^2)) \frac{\partial}{\partial x(\mu^2)}+\nonumber\\&\hspace{.15cm}+2 \gamma_{m}(x(\mu^2)) \frac{\partial}{\partial m_i(\mu^2)}\Bigg\rbrace \mathcal{D}^{J}_2(s)=0\,,
	\end{align}
	where the QCD beta function and the quark mass anomalous dimension ($\gamma_{m}$) are known to the five-loop and can be found in the appendix~\ref{app:mass_run}.\par
	The $SU(3)$ breaking contributions from the Adler function in the determination of quark masses is the difference:
	\begin{align}
		\delta \mathcal{D}^{J,V+A}_{2}(s)&\equiv\mathcal{D}^{J,V+A}_{2,ud}(s)-\mathcal{D}^{J,V+A}_{2,us}(s)
	\end{align}
	where $J=(L+T)/L$ and the analytic expressions can be found in the appendix~\ref{app:dim2adler}. These contributions are used in eq.~\eqref{eq:delRkl} to evaluate the leading order mass correction term $\delta^{(2),kl}_{ud}-\delta^{(2),kl}_{us}$.
	
	The absence of a coefficient $\order{\alpha_s^4}$ for the $``L+T"-$ Adler function induces an additional theoretical uncertainty in the predictions from the perturbation theory. This missing piece can be estimated by $\tilde{d}^{L+T}_{4}\sim (\tilde{d}^{L+T}_{3})^2/\tilde{d}^{L+T}_{2}\approx4067$ and this value is used in the strange quark mass determinations in this article.\par
	The renormalization group running of different coefficients for CIPT and FOPT coefficients can be found in the ref.~\cite{Pich:1998yn,Pich:1999hc}. The RG-summed coefficients can be obtained from appendix~\ref{app:summed_sol} by setting $\left\lbrace n_1,n_2\right\rbrace=\left\lbrace0,2\right\rbrace$.
	
	\subsection{The Dimension-4 contributions to the Adler Function}\label{sec:dim_4_Adler}
	The OPE expansion at dimension-4 involves contributions from perturbative and quark, and gluon condensates \cite{Pich:1999hc,Chetyrkin:1985kn}. However, these contributions  are suppressed by a factor of $(\frac{1}{M_{\tau}^2})^2$, and they have the following form:
	\begin{align}
		\mathcal{D}^{L+T,V/A}_{4,ij}(s)&=\frac{1}{s^2}\sum_{n=0}\tilde{\Omega}^{L+T}_n(s) x(-s)^n\,,\\
		\mathcal{D}^{L,V/A}_{4,ij}(s)&=\frac{1}{M^2_{\tau}\hs s}\bigg\lbrace\frac{3}{2\pi^2}\sum_{n=0}\tilde{\Omega}_n^L(s) x(-s)^n\nonumber\\&\bs-\langle\left(m_i \mp m_j\right)\left(\bar{q}_i q_i \mp \bar{q}_j q_j\right)\rangle\bigg\rbrace\,,
	\end{align}
	where the upper/lower sign corresponds to the V/A component. The $\tilde{\Omega}^{L+T/L}$ coefficients are given by:
	\begin{align}
		\tilde{\Omega}^{L+T}(s)=&\hspace{1mm}\frac{1}{6}\langle G^2\rangle+2\langle m_i \bar{q}_i q_i+m_j\bar{q}_j q_j\rangle \tilde{q}^{L+T}_n\nonumber\\&\pm \frac{8}{3}\langle m_j \bar{q}_i q_i+m_i \bar{q}_j q_j\rangle\tilde{t}^{L+T}_n+\sum_{k}\left\langle m_k \bar{q}_k q_k\right\rangle\nonumber\\&-\frac{3}{\pi^2}\bigg\lbrace\left(m_i^4+m_j^4\right) \tilde{h}^{L+T}_n-m^2_i m^2_j \tilde{g}_n^{L+T}\nonumber\\&\pm\frac{5}{3} m_i m_j\left(m^2_i+m^2_j\right) \tilde{k}_n^{L+T}+\sum_{k}m^4_k \tilde{j}^{L+T}_n\nonumber\\&+2\sum_{k\neq l}m^2_k m^2_l \tilde{u}^{L+T}_n\bigg\rbrace \,, \\
		\tilde{\Omega}^L(s)=&\left(m^2_i+m^2_j\right)\tilde{h}_n^{L}\pm \frac{3}{2}m_i m_j \tilde{k}^L+\sum_{k} m_k^2 \tilde{j}^L_n\,,
	\end{align}
	where $\langle m_i \bar{q}_jq_j\rangle\equiv\langle0\vert m_i \bar{q}_j q_j\vert0\rangle(-\xi^2 s)$, $m_i=m_j(-\xi^2s)$, $\langle G^2\rangle\equiv\langle0\vert G^2\vert0\rangle(-\xi^2 s)$ and $\xi$ is the scale parameter to keep track of the dependence of the renormalization scale.  The RG-evolution of the perturbative coefficients and the condensates can be found in the ref.~\cite{Pich:1999hc}.
	
	The relevant OPE corrections to strange quark mass determination are as follows:
	\begin{align}
		\delta &\mathcal{D}^{L+T}_4(s)\equiv\mathcal{D}^{L+T,V+A}_{4,ud}(s)-\mathcal{D}^{L+T,V+A}_{4,us}(s)\nonumber\\
		&=\frac{-4\delta O_4}{s^2} \sum_{n=0}\tilde{q}^{L+T}_n x(-s)^n+\frac{6}{\pi^2s^2}m_s(-s)^4(1-\epsilon_d^2)\times\nonumber\\&\bs\sum_{n=0}\lbrace\left(1+\epsilon_d^2\right)\tilde{h}^{L+T}_n -\epsilon_u^2\tilde{g}^{L+T}_n\rbrace x(-s)^n\,,
		\label{eq:dim4LTAdler}
	\end{align}
	\begin{align}
		\delta& \mathcal{D}_4^{L}(s)\equiv\mathcal{D}^{L,V+A}_{4,ud}(s)-\mathcal{D}^{L,V+A}_{4,us}(s)\nonumber\\&=\frac{2\delta O_4}{s\mtsq}-\frac{3}{\pi^2 s \mtsq}m_s^4(1-\epsilon_d^2)\nonumber\times\\&\sum_{n=0}\lbrace(1+\epsilon_d^2)(\tilde{h}^{L}_n+\tilde{j}^{L}_n)+\epsilon_u^2(2 \tilde{h}^{L}_n-3\tilde{k}^{L}_n+\tilde{j}_n)\rbrace x(-s)^n\,,
		\label{eq:dim4LAdler}
	\end{align}
	where $\delta O_4=\langle0|m_s \overline{s}s-m_d \overline{d}d|\rangle(-\xi^2 s)$ with $\epsilon_u=m_u/m_s$ and  $\epsilon_u=m_u/m_s$. Using the numerical values
	\begin{align}
		&v_s=0.738\pm0.029   \text{\cite{Albuquerque:2009pr}}\,, \nonumber\\ &f_{\pi}=92.1\pm0.8\MeV\,,\quad m_{\pi}=139.6\MeV \text{\cite{Zyla:2020zbs}}\nonumber\\
		& \epsilon_d=0.053\pm0.002\,, \epsilon_u=0.029\pm0.003\text{\cite{Leutwyler:1996eq}}\,,
	\end{align}
	$\delta O_4$ can be estimated similar to ref.~\cite{Pich:1999hc} as:
	\begin{align}
		\delta O_4&=\left(v_s m_s-m_d\right)\langle0|\overline{d}d|0\rangle\nonumber\\
		&\simeq-\frac{m_s}{2 \hat{m}}\left(v_s-\epsilon_d\right)f_{\pi}^2 m_{\pi}^2\nonumber\\&=-\left(1.54\pm.08\right)\times10^3 \GeV^{-4}\,.
	\end{align}
	
	\section{The behavior of Leading Order Perturbative Mass Corrections in Different Renormalization Schemes}\label{sec:dim_2_behaviour}
	The FOPT and CIPT are the two versions of perturbative theory for the QCD analysis of the $\tau$ decay frequently used in the literature, and it has been further extended by including an RGSPT version of perturbation theory \cite{Abbas:2012py,Ananthanarayan:2016kll}. It has been shown in the ref.~\cite{Abbas:2012py} that the $\delta^{(0)}$ contributions from the RGSPT scheme approach CIPT at higher orders of the perturbation theory, and the corresponding numerical value of the strong coupling constant lies closer to the CIPT value. Similar behavior is also observed in this article for the higher dimensional operators, but with the advantage that the scale dependence for higher moments is under control in the case of RGSPT compared to the FOPT and CIPT. Before moving on to the strange quark mass determination, the convergence behavior of the leading order mass corrections must be analyzed carefully for different schemes. This exercise is performed in the rest of the section. \par
	The leading order mass corrections to moment $\delta R^{kl,2}_{\tau}$ in eq.~\eqref{eq:delRkl} are given by:
	\begin{align}
		\delta R^{kl,D=2}=&24\frac{ m_s(\xi^2\mtsq)^2 }{ M_{\tau}^2}S_{EW} \left(1-\epsilon _d^2\right)\Delta_{kl}(x,\xi)\,,
		\label{eq:rkl2}
	\end{align}
	where,
	\begin{align}
		\Delta_{kl}(x,\xi)\equiv&(\frac{3}{4}\Delta_{kl}^{L+T}(x,\xi)+\frac{1}{4}\Delta_{kl}^{L}(x,\xi))\,,
		\label{eq:Delta}
	\end{align}
	and the $\Delta_{kl}^{J}(x,\xi)$ are the contributions from the Adler functions $\delta\mathcal{D}^{J}$ involving eq.~\eqref{eq:su3adlerLT},\eqref{eq:su3adlerL},\eqref{eq:summed_D2}) evaluated along a contour in the complex plane with the kernels presented in table~\ref{tab:klmoments}. These functions are calculated differently in various schemes explained in the later subsections. \par
	It should be noted that the leading-order mass corrections are presented to remember where perturbative series is truncated in prescription I.
	\subsection{CIPT scheme}\label{sec:CIPT_intro}
	In CIPT, the masses and the strong coupling evolved along the contour in the complex plane by solving the RGE numerically. By construction, it does not suffer from the problem of large logarithm along the contour.
	Following the ref.~\cite{Pich:1998yn,Pich:1999hc,Pich:2020gzz}, dimension-2 contribution to $``L+T"-$component moments can be organized in terms of contour integrals:
	\begin{align}
		\Delta^{L+T}_{kl}(x,\xi)=&-\frac{1}{4\pi i}\sum_{n=0}\tilde{d}^{L+T}_n(\xi)\oint_{|x_c|=1}\frac{dx_c}{x_c^2} \mathcal{F}_{kl}^{L+T}(x_c)\times\nonumber\\& \bs\frac{m_s^2(-\xi^2 M^2_{\tau} x_c)}{m_s^2(M^2_{\tau})} x^n(-\xi^2 M^2_{\tau} x_c)\,,
		\label{eq:delta_lt}
	\end{align}
	and for the longitudinal component:
	\begin{align}
		\Delta^{L}_{kl}(x,\xi)=&\frac{1}{2\pi i}\sum_{n=0}\tilde{d}^{L}_n(\xi)\oint_{|x_c|=1}\frac{dx}{x_c} \mathcal{F}_{kl}^{L}(x_c)\times\nonumber\\&\bs\frac{m_s^2(-\xi^2 M^2_{\tau} x)}{m_s^2(\mtsq)} x^n(-\xi^2 M^2_{\tau} x_c)\,,
		\label{eq:delta_l}
	\end{align}
	The dimension-2 contributions to $\Delta_{kl}^{L+T}$ for $x(\mtsq)=0.3187/\pi$ contributions of different orders are given by:
	\begin{align}     \label{eq:dim2_CI}
		\Delta ^{L+T}_{0,0}&=\{0.7717,0.2198,0.0777,-0.0326,-0.135\}\nonumber\,,\\
		\Delta ^{L+T}_{1,0}&=\{0.9247,0.3324,0.1951,0.0866,-0.0375\}\nonumber\,,\\
		\Delta ^{L+T}_{2,0}&=\{1.0605,0.4410,0.3202,0.2302,0.1019\}\nonumber\,,\\
		\Delta ^{L+T}_{3,0}&=\{1.1883,0.5504,0.4567,0.4021,0.2897\}\nonumber\,,\\
		\Delta ^{L+T}_{4,0}&=\{1.3130,0.6634,0.6073,0.6065,0.5337\}\,.
	\end{align}
	which shows good convergence up to $(2,0)-$moment. The longitudinal contributions are:
	\begin{align}
		\Delta ^L_{0,0}&=\{1.6031,1.1990,1.1583,1.3023,1.6245\}\nonumber\,,\\
		\Delta ^L_{1,0}&=\{1.3832,1.1358,1.1970,1.4642,1.9856\}\nonumber\,,\\
		\Delta ^L_{2,0}&=\{1.2563, 1.1158, 1.2635, 1.6553, 2.4004\}\nonumber\,,\\
		\Delta ^L_{3,0}&=\{1.1783, 1.1204, 1.3494, 1.8740, 2.8757\}\nonumber\,,\\
		\Delta ^L_{4,0}&=\{1.1301, 1.1418, 1.4517, 2.1216, 3.4196\}\,,
		\label{eq:CIPT_L}
	\end{align}
	and we can see that longitudinal contributions show divergent behavior. The total perturbative contributions of dimension-2 is obtained using eq.~\eqref{eq:Delta} are:
	\begin{align}
		\Delta_{0,0}&=\{0.9795, 0.4646, 0.3478, 0.3011, 0.3050\}\nonumber\,,\\
		\Delta_{1,0}&=\{1.0393, 0.5333, 0.4456, 0.4310, 0.4682\}\nonumber\,,\\
		\Delta_{2,0}&=\{1.1094, 0.6097, 0.5560, 0.5865, 0.6765\}\nonumber\,,\\
		\Delta_{3,0}&=\{1.1858, 0.6929, 0.6799, 0.7701, 0.9362\}\nonumber\,,\\
		\Delta_{4,0}&=\{1.2673, 0.7830, 0.8184, 0.9853, 1.2552\}\,.
		\label{eq:Delta_CIPT}
	\end{align}
	It is clear from the eq.~\eqref{eq:Delta_CIPT} that the pathological longitudinal contributions are a restricting factor in getting any reliable determination from CIPT unless we truncate the perturbative series to the minimum term.
	\subsection{FOPT scheme}
	In FOPT, the perturbative series for the Adler function is truncated to a given order in $\alpha_s(\mu)$, and running logarithms are integrated analytically along the contour in the complex energy plane\cite{Pich:1998yn,Beneke:2008ad}. The $\Delta_{kl}^{J}$ for FOPT is evaluated by inserting eq.~\eqref{eq:su3adlerLT1},\eqref{eq:su3adlerL1} in eq.~\eqref{eq:rkl2} and can be written as:
	\begin{align}
		\Delta_{kl}^{J}(x,\xi)=\sum_{i=0}^{4} \sum_{j=0}^{i} x^i(\xi^2 \mtsq)\hs\tilde{d}^{J}_{i,j}H^{kl,J}_{j}(x,\xi) \,,
	\end{align}
	where $H_i^{kl,J} (\xi)$ are evaluated analytically:
	\begin{align}
		H^{kl,L+T}_n(x,\xi)\equiv&\frac{-1}{4\pi i}\oint_{|x_c|=1} \frac{dx_c}{x_c^2}\mathcal{F}^{L+T}_{kl}(x_c)\log^n\left(\frac{-\xi^2}{ x_c}\right)\,,\\
		H^{kl,L}_n(x,\xi)\equiv&	\frac{1}{2\pi i}\oint_{|x_c|=1} \frac{dx_c}{x_c}\mathcal{F}^{L}_{kl}(x_c)\log^n\left(\frac{-\xi^2}{x_c}\right)\,.
	\end{align}
	\par Evaluating the above integrals, the $\Delta_{kl}^{L+T}$ contribution, using FOPT, at different orders of perturbative series is given by:
	\begin{align}
		\Delta ^{L+T}_{0,0}&=\{1.0000,0.4058,0.2575,0.1544,0.0163\}\nonumber\,,\\
		\Delta ^{L+T}_{1,0}&=\{1.0000,0.5072,0.4168,0.3679,0.2971\}\nonumber\,,\\
		\Delta ^{L+T}_{2,0}&=\{1.0000,0.5782,0.5366,0.5414,0.5429\}\nonumber\,,\\
		\Delta ^{L+T}_{3,0}&=\{1.0000,0.6323,0.6330,0.6892,0.7636\}\nonumber\,,\\
		\Delta ^{L+T}_{4,0}&=\{1.0000,0.6758,0.7140,0.8189,0.9654\}\,.
		\label{eq:dim2_FO}
	\end{align}
	$\Delta_{kl}^{L}$ are given by:
	\begin{align}
		\Delta ^L_{0,0}&=\{1.0000,0.9468,1.1319,1.3807,1.7855\}\nonumber\,,\\
		\Delta ^L_{1,0}&=\{0.7500,0.7482,0.9442,1.2183,1.6559\}\nonumber\,,\\
		\Delta ^L_{2,0}&=\{0.6000,0.6229,0.8184,1.1006,1.5520\}\nonumber\,,\\
		\Delta ^L_{3,0}&=\{0.5000,0.5360,0.7271,1.0098,1.4662\}\nonumber\,,\\
		\Delta ^L_{4,0}&=\{0.4286,0.4718,0.6570,0.9371,1.3937\}\,.
		\label{eq:FOPT_L}
	\end{align}
	We can see that the longitudinal piece has a bad convergence in the FOPT scheme. The total contribution $\Delta_{kl}$ is:
	\begin{align}
		\Delta_{0,0}&=\{1.0000, 0.5410, 0.4761, 0.4610, 0.4586\}\nonumber\,,\\
		\Delta_{1,0}&=\{0.9375, 0.5675, 0.5486, 0.5805, 0.6368\}\nonumber\,,\\
		\Delta_{2,0}&=\{0.9000, 0.5894, 0.6071, 0.6812, 0.7952\}\nonumber\,,\\
		\Delta_{3,0}&=\{0.8750, 0.6082, 0.6565, 0.7694, 0.9393\}\nonumber\,,\\
		\Delta_{4,0}&=\{0.8571, 0.6248, 0.6997, 0.8484, 1.0725\}\,.
		\label{eq:Delta_FOPT}
	\end{align}
	We can see that the convergence behavior of dimension-2 contribution in eq.~\eqref{eq:Delta_FOPT} is not very different from the CIPT scheme in eq.~\eqref{eq:Delta_CIPT}.
	\subsection{RGSPT Scheme}
	In Optimal renormalization, masses and coupling are fixed at some renormalization scale, but the RG-summed running logarithms are evolved around the contour. Interestingly, contour integration can be done analytically, similar to FOPT. However, due to the summation of the running logarithms, the resulting perturbative contributions are very much closer to the CIPT numbers and can be seen later in this subsection.\par
	The perturbative series in RGSPT scheme for dimension-2 Adler function has form:
	\begin{align}
		\Delta_{kl}^{J}(x,\xi)=\sum_{i=0}^4 \sum_{n=0}^i \sum_{m=0}^n x^i\left(\xi^2\mtsq\right) \hs\tilde{T}^{J}_{i,n,m} K^{kl,J}_{n,m}(x,\xi)\,,
	\end{align}
	which is obtained by inserting eq.~\eqref{eq:summed_D2} into eq.~\eqref{eq:rkl2} and the corresponding contour integrals $K^{kl,J}_{n,m}(x,\xi)$ have the following form:
	\begin{align}
		K^{kl,L+T}_{n,m}(x,\xi)\equiv&\frac{-1}{4\pi i}\oint_{|x_c|=1} \frac{dx_c}{x_c^2}\mathcal{F}^{L+T}_{kl}(x_c)\times\nonumber\\&\frac{\log^n(1-\beta_0x(\xi^2\mtsq)\log(-\xi^2/ x_c))}{(1-\beta_0 x(\xi^2\mtsq)\log(-\xi^2/ x_c))^m}\,,\\
		K^{kl,L}_{n,m}(x,\xi)\equiv&	\frac{1}{2\pi i}\oint_{|x_c|=1} \frac{dx_c}{x_c}\mathcal{F}^{L}_{kl}(x_c)\times\nonumber\\&\frac{\log^n(1-\beta_0x(\xi^2\mtsq)\log(-\xi^2 /x_c))}{(1-\beta_0 x(\xi^2\mtsq)\log(-\xi^2 /x_c))^m}\,.
	\end{align}
	\par The $\Delta^{L+T}_{i,0}$ contributions for different moments are given by:
	\begin{align}
		\Delta ^{L+T}_{0,0}&=\{0.8878, 0.2307, 0.0799, -0.0328, -0.1561\}\nonumber\,,\\
		\Delta ^{L+T}_{1,0}&=\{0.9990, 0.3690, 0.2263, 0.1220, -0.0176\}\nonumber\,,\\
		\Delta ^{L+T}_{2,0}&=\{1.0885, 0.4931, 0.3736, 0.2980, 0.1691\}\nonumber\,,\\
		\Delta ^{L+T}_{3,0}&=\{1.1652, 0.6095, 0.5246, 0.4957, 0.4043\}\nonumber\,,\\
		\Delta ^{L+T}_{4,0}&=\{1.2336, 0.7212, 0.6806, 0.7153, 0.6891\}\,.
	\end{align}
	and the $\Delta^{L}_{i,0}$ have the form:
	\begin{align}
		\Delta ^L_{0,0}&=\{1.4048, 1.2210, 1.2280, 1.3899, 1.7560\}\nonumber\,,\\
		\Delta ^L_{1,0}&=\{1.1360, 1.1034, 1.2194, 1.5005, 2.0514\}\nonumber\,,\\
		\Delta ^L_{2,0}&=\{0.9687, 1.0302, 1.2287, 1.6169, 2.3536\}\nonumber\,,\\
		\Delta ^L_{3,0}&=\{0.8538, 0.9808, 1.2477, 1.7375, 2.6655\}\nonumber\,,\\
		\Delta ^L_{4,0}&=\{0.7697, 0.9459, 1.2727, 1.8617, 2.9888\}\,,
		\label{eq:RGSPT_L}
	\end{align}
	and the $\Delta(x)$ behave as:
	\begin{align}
		\Delta_{0,0}&=\{1.0171, 0.4783, 0.3669, 0.3229, 0.3219\}\nonumber\,,\\
		\Delta_{1,0}&=\{1.0332, 0.5526, 0.4746, 0.4666, 0.4996\}\nonumber\,,\\
		\Delta_{2,0}&=\{1.0585, 0.6274, 0.5874, 0.6277, 0.7153\}\nonumber\,,\\
		\Delta_{3,0}&=\{1.0874, 0.7023, 0.7054, 0.8061, 0.9696\}\nonumber\,,\\
		\Delta_{4,0}&=\{1.1176, 0.7774, 0.8286, 1.0019, 1.2640\}\,,
		\label{eq:Delta_RGSPT}
	\end{align}
	We can see from the numerical values provided in eq.~\eqref{eq:CIPT_L}, eq.~\eqref{eq:FOPT_L} and eq.~\eqref{eq:RGSPT_L} that the longitudinal contributions have a convergence issue, and it is difficult to get the reliable determinations using them as input. However, the important ingredient in the mass determination is $\Delta_{i,j}$, defined in eq.~\eqref{eq:Delta}. We can see from the numerical values presented in eq.~\eqref{eq:Delta_CIPT}, eq.~\eqref{eq:Delta_FOPT} and eq.~\eqref{eq:Delta_RGSPT} that these inputs can be taken in the mass determination if we truncate the perturbation series to the term which gives minimum contribution to it. This minimum term of the perturbative series is taken as the truncation uncertainty. This prescription has already been advocated in ref.~\cite{Chen:2001qf}, and we have termed this procedure of truncation as \textbf{prescription I}. Another choice is to use all available terms of the perturbation series coefficients of the Adler function, including the estimate for the unknown $\order{\alpha_s^4}$ term of the $``L+T"-$component of the dimension-2 Adler function and termed as \textbf{prescription II}. These prescriptions have some advantages and disadvantages, which will be discussed later.
	\begin{table}
		\begin{center}
			\begin{tabular}{|c|c|c|}
				\hline
				\multirow{1}{*}{\shortstack{Moments\\(k,l)}}&\multicolumn{2}{c|}{$\delta R^{kl}_{\tau}$}\\\cline{2-3}
				&ALEPH &OPAL\\	\hline
				\cline{2-3}
				(0,0)&$0.374 \pm 0.133$&$0.332 \pm 0.10 $\\ \hline
				(1,0)&$0.398 \pm0.077$&$0.326 \pm 0.078 $\\ \hline
				(2,0)&$0.399 \pm 0.053$&$0.340 \pm 0.058$\\ \hline
				(3,0)&$0.396 \pm 0.042$&$0.353 \pm 0.046$\\ \hline
				(4,0)&$0.395 \pm 0.034$&$0.367 \pm 0.037$\\ \hline
			\end{tabular}
		\end{center}
		\caption{Spectral moments from ALEPH \cite{ALEPH:1999uux,Chen:2001qf} and OPAL\cite{OPAL:2004icu}. OPAL moments are calculated using current value of $\vert V_{us}\vert =0.2243\pm0.0008$ quoted in the PDG\cite{Zyla:2020zbs}.}
		\label{tab:rkl_exp}
	\end{table}
	\section{Phenomenological contribution to the longitudinal sector}\label{sec:rev_pheno}
	We can see from the section~\ref{sec:dim_2_behaviour} that although the contributions from the $``L+T"$ part of the dimension-2 has a better convergence for CIPT and RGSPT relative to the FOPT, the longitudinal contributions are forcing us to truncate the higher-order terms. These pathological contributions get enhanced for higher moments and restrict one to use only the leading-order term of the perturbation series. This problem is cured by replacing the longitudinal perturbative series contributions with the phenomenological contributions from the chiral perturbation theory \cite{Gamiz:2002nu,Jamin:2001zq,Jamin:2001zr,Maltman:2001gc,Maltman:2001sv}. These contributions carry significantly less theoretical uncertainty and agree well with the corresponding pQCD results, as shown in ref.~\cite{Gamiz:2002nu}. With these advantages at hand, the strange quark mass determination using the pQCD contribution from the $``L+T"-$component of the Adler function combined with phenomenological contributions for longitudinal contributions in section~\ref{sec:pheno_ms} can be performed. \par
	The relevant quantities of interest for phenomenological contributions to $R^{kl,L}_{ij,V/A}$, the longitudinal component of eq.~\eqref{Rtau_def}, are vector/axial-vector spectral functions $\rho^{V/A}_{ij}(s)$. They are related by:
	\begin{align}
		R^{kl,L}_{ij,V/A}=-24\pi^2\int_{0}^{1}dx_c\left(1-x_c\right)^{2+k}x_c^{l+1} \rho^{V/A,L}_{ij}(\mtsq x_c)\,.
		\label{eq:Rkl_pheno}
	\end{align}
	The pseudoscalar spectral function receives contributions from pion and kaon mass poles and higher resonances in the strange and non-strange channels. We are using the Maltman and Kambor \cite{Maltman:2001sv} parametrization of the pseudoscalar spectral function for the $us$ and $ud$ channels in our analysis, which is given by:
	\begin{align}
		s^2 \rho^{A,L}_{us}(s)=2 f_K^2 m_K^2\delta(s-m_K^2)+\sum_{i=1,2}2 f_i^2 M_i^2 B_i(s)\,.
		\label{eq:rho_pseudoscalar}
	\end{align}
	Here $f_i$ and $M_i$ are the decay constants and masses of resonances, and $B_i(s)$ is the Briet-Wigner resonance function taking the form:
	\begin{align}
		B_i(s)=\frac{1}{\pi} \frac{\Gamma_i M_i}{(s-M_i^2)^2+\Gamma_i^2 M_i^2}\,,
	\end{align}
	where $\Gamma_i $ is the decay width of the resonances. The spectral function for the $ud$ channel is obtained by replacing the kaon terms with the pion in eq.~\eqref{eq:rho_pseudoscalar}. For the resonance contributions to pseudoscalar $ud$ and $us$ channels appearing eq.~\eqref{eq:rho_pseudoscalar}, we have used the following data:
	\begin{table}[H]
		\centering
		\begin{tabular}{|c|c|c|c|c|}
			\hline
			\text{ }&$\pi(1300)$&$\pi(1800)$&$K(1460)$&$K(1800)$\\\hline
			$M_i(\MeV)$& 1300&1810&1482&1830\\\hline
			$\Gamma_i(\MeV)$&400&215&335&250\\\hline
			$f_i(\MeV)$&$2.2\pm0.46$&$0.19\pm0.19$&$21.4\pm2.8$&$4.5\pm 4.5$\\\hline
		\end{tabular}
		\caption{Masses and decay width are taken from PDG \cite{Zyla:2020zbs} and decay constants from ref.~ \cite{Maltman:2001gc}.}
	\end{table}
	The vector component of spectral function receives dominant contributions from the scalar channels $K\pi$,$K\eta$ and $K \eta'$ and the spectral function has the following form\cite{Jamin:2001zr}:
	\begin{align}
		\rho^{V,L}_{uj}(s)=\frac{3\Delta^2_{K\pi}}{32\pi^2}\sum_{i=\{\pi,\eta,\eta'\}}\sigma_{Ki}|F_{Ki}(s)|^2\,,
		\label{eq:rho_vector}
	\end{align}
	where $\Delta_{K\pi}\equiv M_K^2-M_\pi^2$. The phase space factor $\sigma_{Ki}(s)$ are given by:
	\begin{align}
		\sigma_{Ki}(s)=&\theta(s-(M_K+M_i)^2)\times\nonumber\\&\sqrt{\left(1-\frac{\left(M_K+M_i\right)^2}{s}\right)\left(1+\frac{\left(M_K-M_i\right)^2}{s}\right)}\,.
	\end{align}
	The strangeness changing scalar form factors $F_{Ki}(s)$ are defined by:
	\begin{align}
		\langle\Omega|\partial^{\mu}\left(\overline{s}\gamma_{\mu}u |\Omega\rangle \right)\equiv-i \sqrt{\frac{3}{2}} \Delta_{K\pi} F_{K \pi}(s)\,,
	\end{align}
	and can be found in ref.~\cite{Jamin:2001zq}. Detailed discussion on the application of these form factors in the extraction of strange quark mass can be found in the refs.~\cite{Jamin:2001zr,Gamiz:2002nu}.
	\section{Strange quark mass determination from pQCD}\label{sec:ms_pert}
	The strange quark mass determination in this section is based on the method used in refs.~\cite{Pich:1999hc,Chen:2001qf,Ananthanarayan:2016kll}. In addition, we have employed different schemes to perform the comparative study. The strange quark mass determination from hadronic $\tau$ decays using RGSPT has been performed ref.~\cite{Ananthanarayan:2016kll}. However, the uncertainties coming from the truncation of perturbative series and the scale dependence of strange quark masses were neglected. We have improved the previous determination using pQCD inputs by including these uncertainties and the determinations made in the two prescriptions mentioned in section~\ref{sec:dim_2_behaviour}.\par
	It should be noted that the higher dimensional OPE contributions ($d>4$) to the Adler functions, which are numerically small \cite{Pich:1999hc} and not considered in this analysis. The strange quark mass is determined by supplying experimental and theoretical inputs to eq.~\eqref{eq:delRkl}. The RHS of the equation is provided with theory inputs from dimension-2 contributions eqs.~\eqref{eq:su3adlerLT}, \eqref{eq:su3adlerL} and dimension-4 with eqs.~\eqref{eq:dim4LTAdler},\eqref{eq:dim4LAdler}. These quantities are evaluated along the complex contour in different schemes, as explained in section~\ref{sec:dim_2_behaviour}. We present our weighted averaged determinations for $m_s(\mtsq)$ from different moments in the table~\ref{tab:mspertweighted} for different schemes. The details of various sources of uncertainty in the two prescriptions are presented in appendix~\ref{app:pQCD_mass}.\par
	We can see from table~\ref{tab:mspertweighted} that the strange quark mass determination from different schemes agrees with each other within uncertainty. It is also evident from the tables presented in appendix~\ref{app:pQCD_mass} that the uncertainties in the final strange quark mass are higher in prescription I than in prescription II mainly due to the truncation of the perturbative series. We also emphasize that the systematic comparison of the behavior of perturbative series in different schemes can only be made in prescription II, where the same order information is used. The RGSPT provides better control over the theoretical uncertainty by minimizing the renormalization scale dependence. The scale dependence of the strange quark mass for various moments is shown in the figure~\ref{fig:ms_scdep}, plotted using prescription II. These plots indicate that the strange mass from the RGSPT scheme is stable for a wider range of scale variations for the moments under consideration. It should be noted that the uncertainties associated with renormalization scale dependence are included only in the range $\xi\in\left[0.75,2.0\right]$ in the strange mass determination in table~\ref{tab:mspertweighted}. \par It should be noted that the poor convergence of the longitudinal contributions restricts this method to be applicable in the lower energies $s_0<\mtsq$. Additional uncertainties in the determinations of $m_s$ arise due to the variations of the upper limit of the moment in the integral $s_0$ defined in eq.~\eqref{eq:Rratio}. These are estimated using the phenomenological determination, discussed in the next section, and are also included in the table~\ref{tab:mspertweighted}. Further details on the numerical uncertainties using pQCD inputs can be found in the appendix~\ref{app:pQCD_mass}.
	\begin{widetext}
		
		\begin{figure}[H]
			\begin{subfigure}{.24\textwidth}
				\centering
				\includegraphics[width=\linewidth]{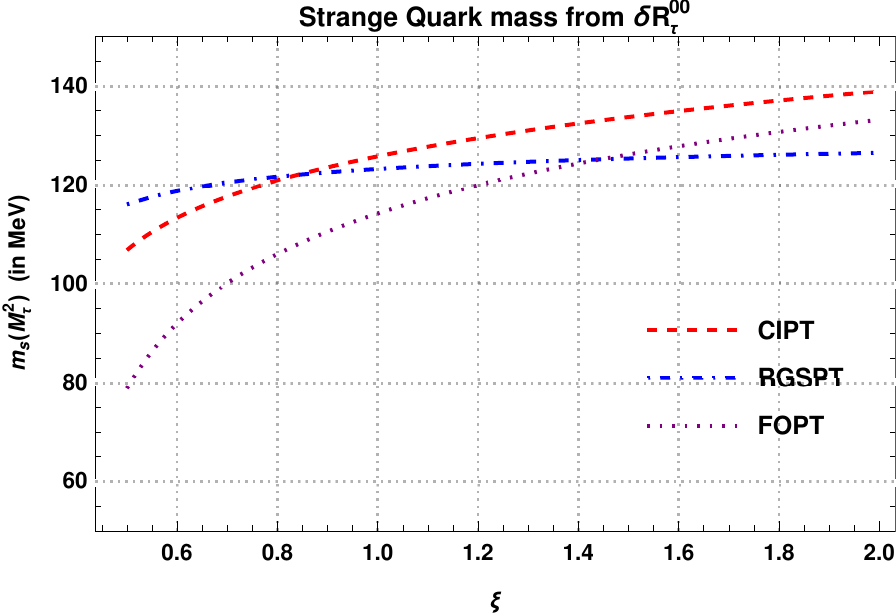}
			\end{subfigure}
			\begin{subfigure}{.24\textwidth}
				\centering
				\includegraphics[width=\linewidth]{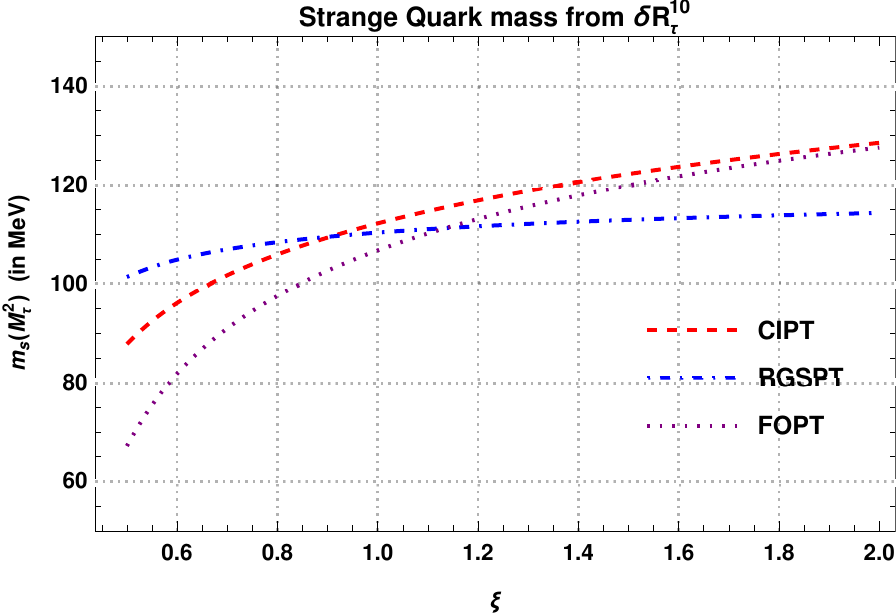}
			\end{subfigure}
			\begin{subfigure}{.24\textwidth}
				\centering
				\includegraphics[width=\linewidth]{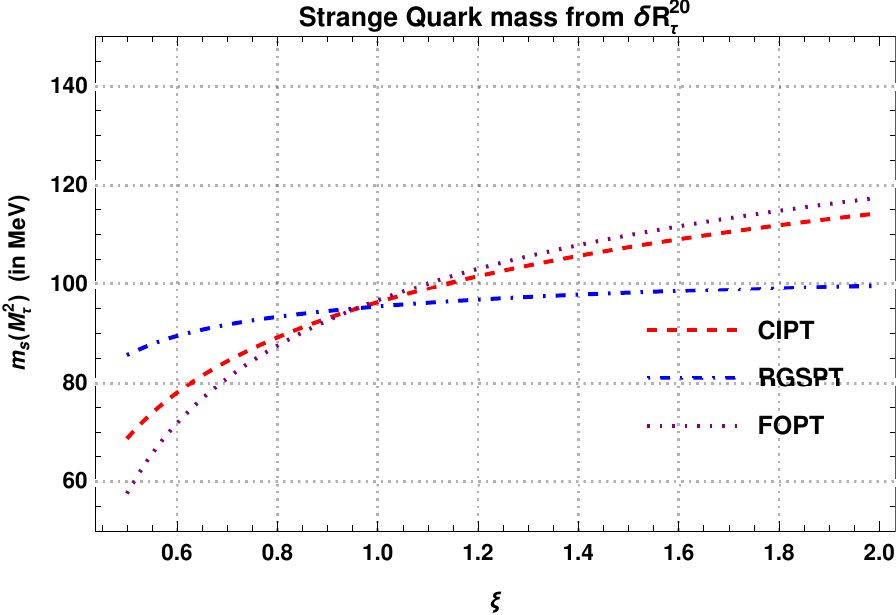}
			\end{subfigure}
			\begin{subfigure}{.24\textwidth}
				\centering
				\includegraphics[width=\linewidth]{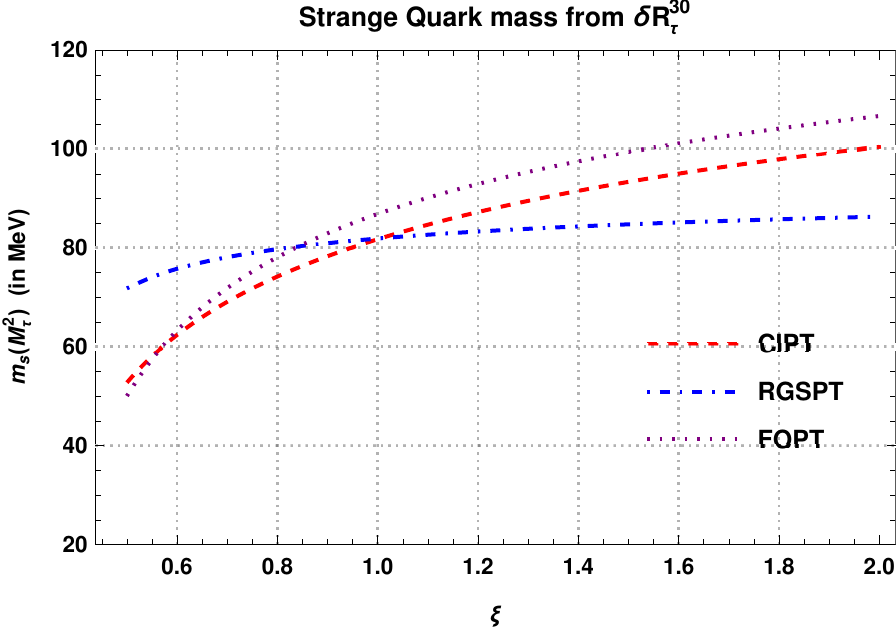}
			\end{subfigure}
			\caption{ The scale variation of the strange quark mass in different perturbative schemes using pQCD inputs.}
			\label{fig:ms_scdep}
		\end{figure}
		%
		%
		
		\begin{table}[H]
			\centering
			\begin{tabular}{|c|c|c|c|c|}
				\hline
				Perturbative Scheme&\multicolumn{2}{c|}{$m_s(\mtsq)$ using prescription I (in $\MeV$)}&\multicolumn{2}{c|}{$m_s(\mtsq)$ using prescription II (in $\MeV$)} \\
				\cline{2-3}\cline{4-5}
				\text{}&ALEPH&OPAL&ALEPH&OPAL\\ \hline
				CIPT& $117.7\pm28.5$&$105.9\pm27.5$&$93.2\pm24.1$&$84.1\pm22.7$\\ \hline
				FOPT&$129.3\pm33.5$&$116.2\pm29.2$&$94.2\pm25.4$&$85.2\pm23.9$\\ \hline
				RGSPT&$120.2\pm23.4$&$107.7\pm25.1$&$89.4\pm16.4$&$80.1\pm15.8$\\ \hline
			\end{tabular}
			\caption{ Weighted average of the strange quark mass in different perturbative schemes. }
			\label{tab:mspertweighted}
		\end{table}

	\end{widetext}

	\section{Strange quark mass determination using phenomenological inputs}\label{sec:pheno_ms}
	The determination of the strange quark mass in this section is similar to the one used in section~\ref{sec:ms_pert}, but now the longitudinal Adler function is replaced with the phenomenologically parameterized contributions, as discussed in section~\ref{sec:rev_pheno}. It should be noted that the $``L+T"$ component of the Adler function at dimension-2 is known to $\order{\alpha_s^3}$ and we do not use its estimate for the $\order{\alpha_s^4}$ coefficient in the determination of $m_s(\mtsq)$ using phenomenological inputs. Contributions from the last known term of the perturbation series of the Adler function are taken as the total truncation uncertainty, similar to that of the previous section. \par
	Following the discussions of section~\ref{sec:rev_pheno}, we now have all the necessary ingredients for the strange quark mass determination. Using the transverse contributions used in section~\ref{sec:ms_pert} and combining them with the input from section~\ref{sec:rev_pheno}, we determine the strange quark in different schemes. We present our result for the weighted average in the table~\ref{tab:msphenoweighted2} and the further details of the determinations from moments in the appendix~\ref{app:pheno_mass}. The scale dependence in the $m_s(\mtsq)$ is presented in the figure~\ref{fig:ms_scdeppheno} using prescription II. As observed in the previous section, the strange quark mass determinations from the RGSPT scheme are stable over the wider range of scale variation for moments under consideration. The determination of the $m_s$ from the traditional spectral moments is sensitive to the variation of the $s_0$. A typical 5\% variation of the $s_0$ from $\mtsq$ in the range $s_0\in\left[3,\mtsq\right]$ induces variations of $\sim 6-13\%$ in the $m_s$ determinations from moments using the OPAL data. Unfortunately, such variations can not be calculated for the ALEPH moments as the strange spectral function is not publicly available. These uncertainties are estimated from the determinations using the OPAL data.
	\begin{widetext}
		
		\begin{table}[H]
			\centering
			\begin{tabular}{|c|c|c|c|c|}
				\hline
				Perturbative Scheme&\multicolumn{2}{c|}{$m_s(\mtsq)$ using prescription I (in $\MeV$)}&\multicolumn{2}{c|}{$m_s(\mtsq)$ using prescription II (in $\MeV$)} \\
				\cline{2-3}\cline{4-5}
				\text{}&ALEPH&OPAL&ALEPH&OPAL\\ \hline
				CIPT& $123.3\pm22.3$&$106.3\pm21.5$&$125.1\pm25.1$&$107.5\pm23.9$\\ \hline
				FOPT&$136.6\pm35.0$&$119.5\pm35.4$&$115.8\pm30.1$&$101.6\pm28.3$\\ \hline
				RGSPT&$123.1\pm21.1$&$107.0\pm21.2$&$117.7\pm20.1$&$102.0\pm19.5$\\ \hline
			\end{tabular}
			\caption{The weighted average of strange quark mass in the different perturbative schemes. Phenomenological inputs for the longitudinal contributions are used.}
			\label{tab:msphenoweighted2}
		\end{table}

		\begin{figure}[H]
			\begin{subfigure}{.24\textwidth}
				\centering
				\includegraphics[width=\linewidth]{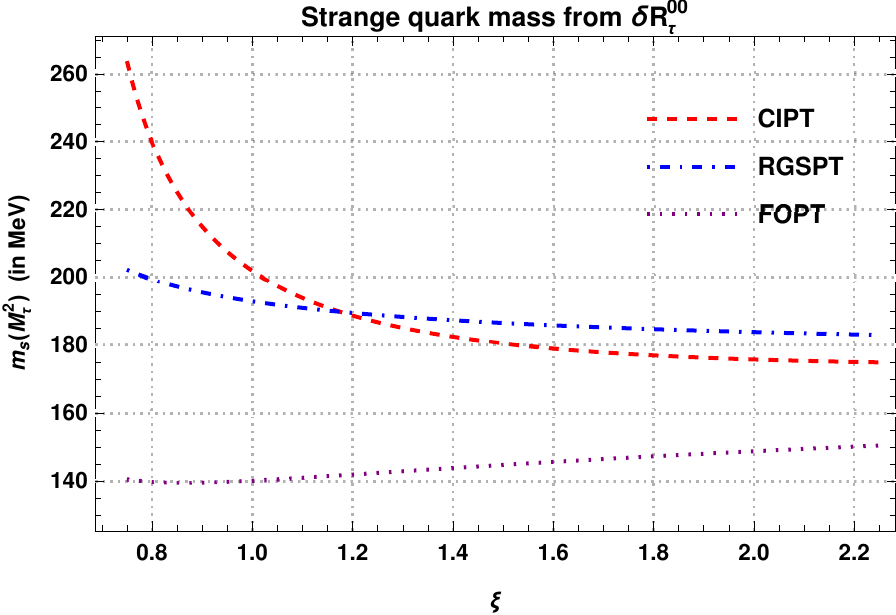}
			\end{subfigure}
			\begin{subfigure}{.24\textwidth}
				\centering
				\includegraphics[width=\linewidth]{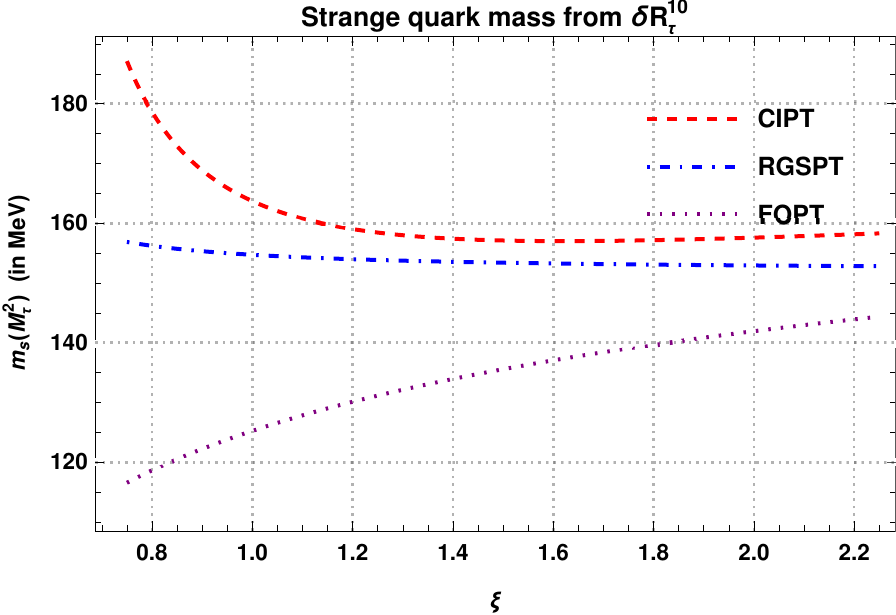}
			\end{subfigure}
			\begin{subfigure}{.24\textwidth}
				\centering
				\includegraphics[width=\linewidth]{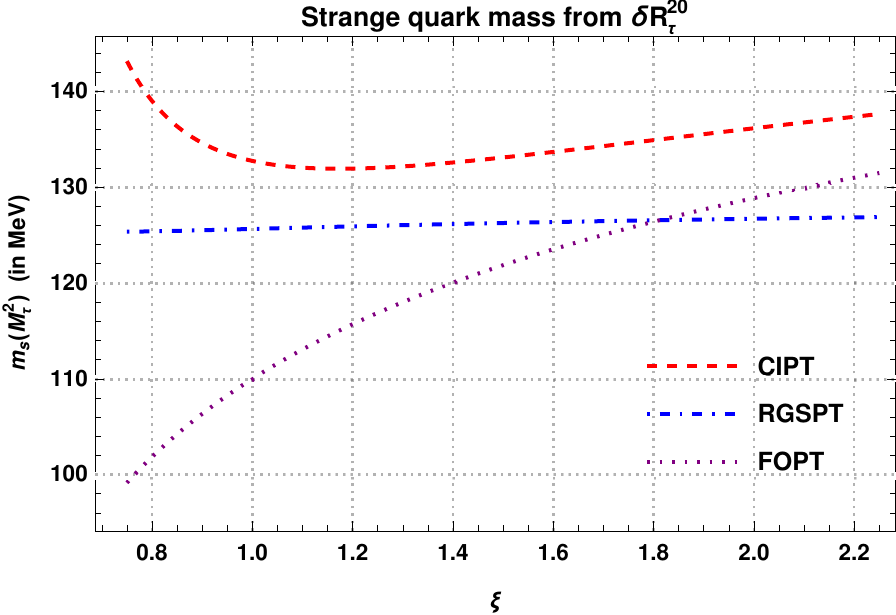}
			\end{subfigure}
			\begin{subfigure}{.24\textwidth}
				\centering
				\includegraphics[width=\linewidth]{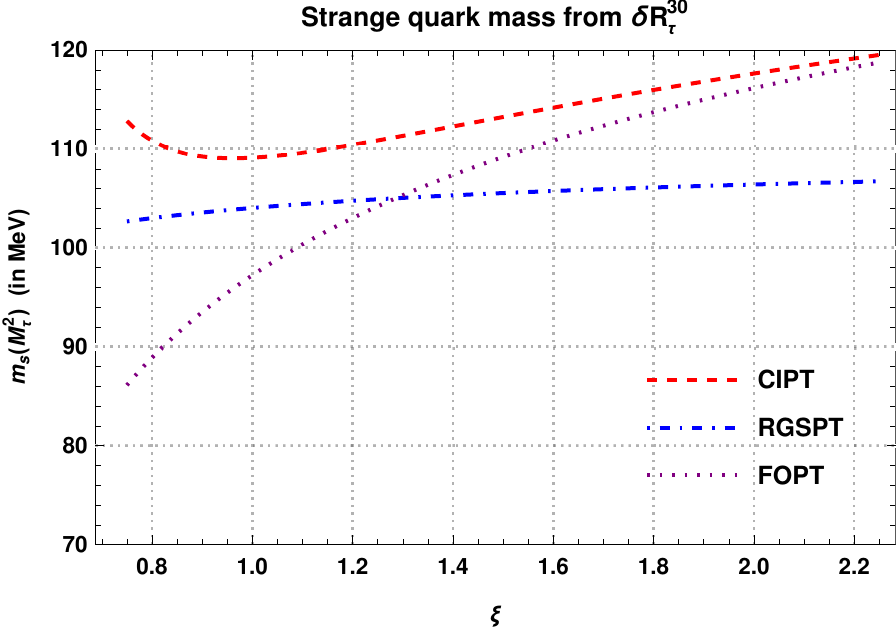}
			\end{subfigure}
			\caption{The scale variation of the strange quark mass obtained from phenomenological inputs in different perturbative schemes.}
			\label{fig:ms_scdeppheno}
		\end{figure}

	\end{widetext}
	
	\section{Determination of \texorpdfstring{$\vert V_{us}\vert$}{}}\label{sec:Vusextraction}
	The data on strange and non-strange spectral moments for the hadronic $\tau$ decay provided by ALEPH \cite{ALEPH:2005qgp,Davier:2008sk,Davier:2013sfa}, HFLAV\cite{HFLAV:2019otj}, and OPAL \cite{OPAL:2004icu} collaborations can be used to determine the CKM matrix element $\vert V_{us}\vert$. These experimental moments along with the theoretical moments calculated with strange quark mass as input from other sources can be used to determine $\vert V_{us}\vert$ using the following relation:
	\begin{align}\label{eq:vus_calc}
		\vert V_{us}\vert=\sqrt{\frac{R_{\tau,S}^{kl}}{R^{kl}_{\tau,V+A}/|V_{ud}|^2-\delta R^{kl}_{\tau,th}}}\,,
	\end{align}
	where $R^{kl}_{\tau,S}$ and $R^{kl}_{\tau,V+A}$ are experimental inputs and $\delta R^{kl}_{\tau,th}$ is the theory input, in which $m_s(2\GeV)$ is taken as an external input. This method has already been used previously in refs.~\cite{Maltman:2007ic,Gamiz:2004ar,Gamiz:2006xx,Gamiz:2004gh,Gamiz:2005gh,Gamiz:2002nu}, and it has been observed that the uncertainties are dominated by the experimental data available for the strange component. An additional source of uncertainties is pointed out in ref.~\cite{Maltman:2007ic} due to $s_0$ variations which can be cured using a different analysis based on the non-spectral weight functions. However, we have restricted this analysis only to the traditional weight functions.\par
	Using $m_s(2\GeV)=93\pm11\MeV$\cite{Zyla:2020zbs} as an external input and ALEPH data \cite{ALEPH:2005qgp,Davier:2008sk,Davier:2013sfa}, we have presented our determination for $\vert V_{us}\vert$ in table~\ref{tab:Vus_Aleph1}.

	\begin{table}
		\centering
		\begin{tabular}{|c|c|c|}
			\hline
			Scheme&\multicolumn{2}{c|}{$\vert V_{us}\vert$}\\
			\cline{2-3}
			\text{}&pQCD inputs&Phenomenological inputs\\
			\hline
			CIPT&$0.2174\pm0.0045$&$0.2168\pm0.0044$\\\hline
			FOPT&$0.2183\pm 0.0055$&$0.2179\pm0.0055$\\\hline
			RGSPT&$0.2178\pm0.0046$&$0.2170\pm0.0045$\\
			\hline
		\end{tabular}
		\caption{$\vert V_{us}\vert$ from ALEPH data in different perturbative scheme using $\delta R^{kl}_{\tau,th} $ from section~\ref{sec:ms_pert} and section~\ref{sec:pheno_ms}.}
		\label{tab:Vus_Aleph1}
	\end{table}

	The latest branching fraction of hadronic $\tau$ decays into non-strange and slightly more precise strange component from HFLAV\cite{HFLAV:2019otj} can also be used to get more precise determination of $\vert V_{us}\vert$ from this method. The results for different schemes are presented in table \ref{tab:Vus_HFLAV1}.\par   The uncertainties shown in these tables are dominated by those coming from the variation of $s_0\in\left[2.5,\mtsq\right]$ and experimental uncertainty in strange $R_{\tau,S}$ contributions. It should be noted that uncertainties coming from the variation of the $s_0$ in table~\ref{tab:Vus_Aleph1} and table~\ref{tab:Vus_HFLAV1} are calculated using the experimental data on the spectral function from ref.~\cite{OPAL:2004icu,Davier:2013sfa}.

	\begin{table}
		\centering
		\begin{tabular}{|c|c|c|}
			\hline
			Scheme&\multicolumn{2}{c|}{$\vert V_{us}\vert$}\\
			\cline{2-3}
			\text{}&pQCD inputs&Phenomenological inputs\\
			\hline
			CIPT&$0.2195\pm0.0047$&$0.2189\pm0.0044$\\\hline
			FOPT&$0.2205\pm 0.0043$&$0.2200\pm0.0037$\\\hline
			RGSPT&$0.2199\pm0.0046$&$0.2191\pm0.0043$\\
			\hline
		\end{tabular}
		\caption{$\vert V_{us}\vert$ from HFLAV data in different perturbative scheme using $\delta R^{kl}_{\tau,th} $ from section~\ref{sec:ms_pert} and section~\ref{sec:pheno_ms}.}
		\label{tab:Vus_HFLAV1}
	\end{table}

	The $\vert V_{us}\vert$ determinations from ALEPH \cite{ALEPH:2005qgp,Davier:2008sk,Davier:2013sfa} and  HFLAV\cite{HFLAV:2019otj} are based on $(0,0)-$moment. A detailed analysis for higher moments can be performed using the OPAL\cite{OPAL:2004icu} data, where $(k,0)$ moments for $k=0,\cdots,4$ are also available. These moments are correlated and their correlation should also be considered in the full analysis. Given the large uncertainties in their strange components and unknown precise higher dimensional OPE corrections, we have neglected these correlations among various moments in our determinations from this data.\par
	Using the strange and non-strange moments from OPAL and the theoretical inputs for $\delta R^{kl}_{\tau,th} $ from Section~\ref{sec:ms_pert}, we present the weighted average of the determination $\vert V_{us}\vert$ from prescription I and prescription II in the table~\ref{tab:Vus_weightedmean12}. Details of the $\vert V_{us}\vert$ from moments along with the sources of uncertainties can be found in table~\ref{tab:VusAllpert1_Opal} and table~\ref{tab:VusAll2pert_Opal}, respectively. We can observe from these tables that the RGSPT is slightly more sensitive to the strange quark mass taken as input, but the overall theory uncertainty coming in this scheme is lesser than CIPT and FOPT in prescription II. We can also see that the divergent nature of the longitudinal component is still an issue causing a large theoretical uncertainty dominating in the higher moments in prescription II in the table~\ref{tab:VusAll2pert_Opal}.\par
	These shortcomings are slightly improved in the phenomenological determination and can be seen in table~\ref{tab:VusAll1pheno_Opal} and table~\ref{tab:VusAll2pheno_Opal}. Again, these determinations suffer from large $s_0$ and  theoretical uncertainties, especially those coming from the strange quark mass in the higher moments. It is worth emphasizing that prescription I reduces the dependence on the spectral moments in the  $\vert V_{us}\vert$ determination. The weighted average of these results for $\vert V_{us}\vert$ are presented in the table~\ref{tab:Vus_weightedmean12}. The RGSPT scheme is slightly more sensitive to the variation of the $s_0$, which, compared to CIPT, dominates in the final average presented in the table~\ref{tab:VusAll1pheno_Opal} and can be seen in the table presented in appendix~\ref{app:vus}. 
	\begin{widetext}
		
		\begin{table}[H]
			\centering
			\begin{tabular}{|c|c|c|c|c|c|}
				\hline
				Perturbative scheme&\multicolumn{2}{c|}{$\vert V_{us}\vert$ from prescription I}&\multicolumn{2}{c|}{$\vert V_{us}\vert$ from prescription II}\\
				\cline{2-3}\cline{4-5}
				\text{}&pQCD inputs&Phenomenological inputs&pQCD inputs&Phenomenological inputs\\
				\hline
				CIPT&$0.2220\pm0.0050$&$0.2212\pm0.0047$&$0.2232\pm0.0048$&$0.2212\pm0.0045$\\ \hline
				FOPT&$0.2212\pm 0.0059$&$0.2220\pm0.0054$&$0.2240\pm 0.0059$&$0.2224\pm0.0054$\\ \hline
				RGSPT&$0.2222\pm0.0051$&$0.2215\pm0.0048$&$0.2238\pm0.0049$&$0.2215\pm0.0047$\\ \hline
			\end{tabular}
			\caption{The weighted average of the determination of $\vert V_{us}\vert$ from the OPAL data in the different perturbative schemes.}
			\label{tab:Vus_weightedmean12}
		\end{table}
		
	\end{widetext}
	\section{Joint \texorpdfstring{$m_s$}{} and \texorpdfstring{$\vert V_{us}\vert$}{} determination}\label{sec:joint_msVus}
	The experimental moments provided by the OPAL collaboration in ref.~\cite{OPAL:2004icu} can be used for the joint extraction of $m_s$ and $\vert V_{us}\vert$. It should be noted that the moments provided in the ref.~\cite{OPAL:2004icu} are correlated, and a proper analysis will require their correlations to be taken into account. Given the uncertainties present in the data, we are disregarding these correlations and restrict ourselves to simplified analysis .\par Using the phenomenological parametrization for the longitudinal contributions and perturbative $``L+T"-$component from section~\ref{sec:rev_pheno} and section~\ref{sec:OPE_contributions}, we fit the $m_s(2\GeV)$ and $\vert V_{us}\vert$ to eq.~\eqref{eq:vus_calc} for moments $(k,0)$ with $k=0,1,\cdots,4$. The central values of the joint fit are presented in table~\ref{tab:msVusAll_Opal1}. These joint fits give smaller values for $m_s(2\GeV)$ and $\vert V_{us}\vert$ compared to the PDG average \cite{Zyla:2020zbs}, but very close to the findings of Gamiz et al. \cite{Gamiz:2004ar} for CIPT and RGSPT.
	
	\begin{table}
		\centering
		\begin{tabular}{|c|c|c|c|c|}
			\hline
			Scheme&\multicolumn{4}{c|}{Phenomenological inputs}\\
			\cline{2-5}
			\text{}&\multicolumn{2}{c|}{prescription I}&\multicolumn{2}{c|}{prescription II}\\
			\cline{2-3}\cline{4-5}
			\text{}&$m_s(2\MeV)$&$\vert V_{us}\vert$&$m_s(2\MeV)$&$\vert V_{us}\vert$\\
			\hline
			CIPT&75&0.2199&75&0.2199\\\hline
			FOPT&46&0.2227&46&0.2227\\\hline
			RGSPT&73&0.2199&73&0.2199\\
			\hline
		\end{tabular}
		\caption{Joint determination of $m_s(2\GeV)$ and $\vert V_{us}\vert$ from OPAL data with moment~(k,0) with k={0,1,2,3,4} in different perturbative scheme from pQCD inputs as well as using phenomenological inputs.}
		\label{tab:msVusAll_Opal1}
	\end{table}
	
	\section{Summary and conclusion}\label{sec:summary}
	The hadronic $\tau$ decays are important ingredients for extracting various QCD parameters. We have used perturbative schemes CIPT, FOPT, and RGSPT in the extraction of $m_s$, $\vert V_{us}\vert$, and their joint determinations from the experimental inputs available from ALEPH \cite{ALEPH:2005qgp,Davier:2008sk,Davier:2013sfa}, HFLAV\cite{HFLAV:2019otj}, and OPAL \cite{OPAL:2004icu} moments of hadronic $\tau$ decays. To reach the goal, we first calculate the RGSPT coefficients for the dimension-4 operator and use them for determinations of $m_s$ and $\vert V_{us}\vert$. Dimension-6 OPE corrections are known to NLO and their RG-improvement is discussed in ref.~\cite{Lanin:1986zs,Adam:1993uu,Boito:2015joa}. Four quark condensates present in these contributions are estimated using the vacuum saturation approximation~\cite{Shifman:1978bw} and found to be numerically very small and are not considered in this article. Higher dimensional OPE corrections are not fully known and are neglected in this article. \par
	The moments calculated using the perturbation theory suffer from convergence issues, so we have employed two prescriptions. The central values of the strange quark mass determinations from prescription I are less spread for different moments than in prescription II. The theoretical uncertainties arising from truncation and scale dependence dominate for higher moments in the prescription I for CIPT and FOPT. However, RGSPT has better control over the scale dependence for a wider range of scale variation, even for higher moments as shown in figure~\ref{fig:ms_scdep} and figure~\ref{fig:ms_scdeppheno}. This improvement results in a more precise determination in RGSPT compared to FOPT and CIPT schemes.\par
	The important results of this article for the $m_s(\mtsq)$ determination are presented in table~\eqref{tab:mspertweighted},\eqref{tab:msphenoweighted2} and for $\vert V_{us}\vert$ determinations in tables~\eqref{tab:Vus_weightedmean12},~\eqref{tab:Vus_Aleph1}~and \eqref{tab:Vus_HFLAV1}. The joint $m_s$ and $\vert V_{us}\vert$ determination results are presented in table~\ref{tab:msVusAll_Opal1}. It should be noted that the ALEPH moments used in the $m_s$ determinations in this article are based on the old $\vert V_{us}\vert$ calculated in ref.~\cite{Chen:2001qf}. The strange quark mass determinations from the moments are very sensitive to the value of $\vert V_{us}\vert$, and hence we do not consider them in the final average. However, the experimental data for the strange and non-strange moments provided by the OPAL collaboration in ref.~\cite{OPAL:2004icu} with the current value of $\vert V_{us}\vert=0.2245\pm0.0008$\cite{Zyla:2020zbs} as an input can be used to provide the most updated determination of the strange quark mass.\par We give our final determination for $m_s(\mtsq)$, which comes from the weighted average of strange quark mass determination using RGSPT scheme from table~\ref{tab:msphenoweighted2}:
	\begin{align}
		m_s(\mtsq)&=102.0\pm19.5\MeV\bs\text{(OPAL,RGSPT)}\,,
	\end{align}
	which corresponds to the strange quark mass at $2\GeV$ :
	\begin{align}
		m_s(2\GeV)&=98\pm19\MeV\,
	\end{align}
	\par
	Using ALEPH\cite{Davier:2008sk,Davier:2013sfa} moment, the $\vert V_{us}\vert$ determinations along with their deviation from PDG\cite{Zyla:2020zbs} ( $\vert V_{us}\vert =0.2243\pm0.0008$) and CKM unitarity fit value ($\vert V_{us}\vert=0.2277\pm0.0013$) are:
	\begin{align}
		\vert V_{us}\vert&=0.2168\pm0.0044\quad(1.7\hspace{.5mm}\sigma,2.6\hspace{.5mm}\sigma)&\hs(\text{for CIPT}) \,,\\
		\vert V_{us}\vert&=0.2170\pm0.0045\quad(1.6\hspace{.5mm}\sigma,2.3\hspace{.5mm}\sigma)&\hs(\text{for RGSPT})\,,
	\end{align}
	and from HFLAV\cite{HFLAV:2019otj}:
	\begin{align}
		\vert V_{us}\vert&=0.2189\pm0.0044\quad(1.2\hspace{.5mm}\sigma,1.9\hspace{.5mm}\sigma)&\hs(\text{for CIPT}) \,,\\
		\vert V_{us}\vert&=0.2191\pm0.0043\quad(1.2\hspace{.5mm}\sigma,1.9\hspace{.5mm}\sigma)&\hs(\text{for RGSPT})\,.
	\end{align}
	The weighted average of the $\vert V_{us}\vert$ determinations from OPAL\cite{OPAL:2004icu} using phenomenological inputs is presented in table~\ref{tab:Vus_weightedmean12}. The most precise determinations for $\vert V_{us}\vert$ from this table come from CIPT and RGSPT:
	\begin{align}
		\vert V_{us}\vert&=0.2212\pm0.0047, 0.2212\pm0.0045&\hs(\text{for CIPT}) \,,\\
		\vert V_{us}\vert&=0.2215\pm0.0048,0.2215\pm0.0047&\hs(\text{for RGSPT})\,.
	\end{align}
	The mean values of determinations in these schemes are:
	\begin{align}
		\vert V_{us}\vert&=0.2212\pm 0.0045\quad(0.7\hspace{.5mm}\sigma,1.4\hspace{.5mm}\sigma)&\hs(\text{for CIPT}) \,,\\
		\vert V_{us}\vert&=0.2215\pm 0.0047\quad(0.6\hspace{.5mm}\sigma,1.3\hspace{.5mm}\sigma)&\hs(\text{for RGSPT}) \,.
	\end{align}\par
	We give our final determinations by weighted average of these results as:
	\begin{align}
		\vert V_{us}\vert&=0.2189\pm0.0044\quad(1.2\hspace{.5mm}\sigma,1.9\hspace{.5mm}\sigma)&\hs(\text{for CIPT}) \,,\\
		\vert V_{us}\vert&=0.2191\pm0.0043\quad(1.2\hspace{.5mm}\sigma,1.9\hspace{.5mm}\sigma)&\hs(\text{for RGSPT})\,.
	\end{align}
	The values obtained for $\vert V_{us}\vert$ using OPAL data agrees with the PDG average within uncertainties. However, the $\vert V_
	{us}\vert$ determination from ALEPH \cite{ALEPH:2005qgp,Davier:2008sk,Davier:2013sfa} and  HFLAV\cite{HFLAV:2019otj} are more than $1.2\sigma$ and $1.9\sigma$ away from the PDG\cite{Zyla:2020zbs} average and CKM unitarity fit value. It should be noted that the PDG average is already in tension with $2.2\hspace{.5mm}\sigma$ with the CKM unitarity.\par
	The dependence of our determinations on the choice of the moments and their correlation is not considered in this article, and we expect that these can be further improved using non-spectral weights used by Maltman et al. in ref.~\cite{Kambor:2000dj,Maltman:2006st,Hudspith:2017vew}.

	\section{Acknowledgments}
	We thank Prof. Diogo Boito for providing help with the experimental data on the spectral functions. BA is partly supported by the MSIL Chair of the Division of Physical and Mathematical Sciences, Indian Institute of Science. DD would like to thank the DST, Govt. of India for the INSPIRE Faculty Award (grant no IFA16-PH170). DD also thanks the Institute for Theoretical Physics III, University of Stuttgart for kind hospitality during various stages of the work. AK thanks Shiuli Chatterjee, Rhitaja Sengupta, and Prasad Hegde for their valuable discussions. AK is supported by a fellowship from the Ministry of Human Resources Development, Government of India.
	\appendix
	
	\newpage
	
	\begin{widetext}
		
		\section{Running of the Strong Coupling and the Quark Masses in the pQCD} \label{app:mass_run}
		
		The running of strong coupling and the quark masses are computed by solving the following differential equations:
		\begin{align}
			\mu^2\frac{d}{d\mu^2}x(\mu^2)=\beta(x(\mu^2))=-\sum_i x^{i+2}(\mu^2) \beta_i\,, \quad
			\mu^2 \frac{d}{d\mu^2}m_i(\mu^2)\equiv&\hspace{2mm}m_i(\mu^2) \hspace{.4mm}\gamma_{m} =	-m_i(\mu^2)\sum_{i}\gamma_i \hspace{.4mm}x^{i+1}(\mu^2)\,.
			\label{anomalous_dim}
		\end{align}
		\par The series solutions for the running of strong coupling and the quark masses relevant for contour integration in the complex plane using FOPT are:%
		\begin{align}
			x(q^2)= x \Bigg\lbrace1&+ x \beta _0 L+x^2 \left(\beta _1 L+\beta _0^2 L^2\right)+x^3 \left(\beta _2 L+\frac{5}{2} \beta _1 \beta _0 L^2+\beta _0^3 L^3\right)\nonumber\\&+x^4 \Big(\beta _3 L+\big(\frac{3 \beta _1^2}{2}+3 \beta _0 \beta _2\big) L^2+\frac{13}{3} \beta _1 \beta _0^2 L^3+\beta _0^4 L^4\Big)\nonumber\\&+x^5\Big( \beta _4 L+\big(\frac{7 \beta _1 \beta _2}{2}+\frac{7 \beta _0 \beta _3}{2}\big) L^2+\big(6 \beta _2 \beta _0^2+\frac{35}{6} \beta _1^2 \beta _0\big) L^3+\frac{77}{12} \beta _1 \beta _0^3 L^4+\beta _0^5 L^5\Big) \Bigg\rbrace+ \order{\alpha_s^7}\,,
			\label{eq:alphasmu}\\
			m_i(q^2)=m0_i\Bigg\lbrace1&+x \gamma _0 L+x^2 \left(\gamma _1 L+\frac{1}{2} \gamma _0 L^2 \left(\beta _0+\gamma _0\right)\right)\nonumber\\&+x^3 \Big(\gamma _2 L+L^2 \left(\frac{\beta _1 \gamma _0}{2}+\gamma _1 \left(\beta _0+\gamma _0\right)\right)+\frac{1}{6} \gamma _0 L^3 \left(\beta _0+\gamma _0\right) \left(2 \beta _0+\gamma _0\right)\Big)\nonumber\\&+x^4 \Bigg(\gamma _3 L+L^2 \left(\beta _1 \gamma _1+\frac{\beta _2 \gamma _0}{2}+\frac{3 \beta _0 \gamma _2}{2}+\frac{\gamma _1^2}{2}+\gamma _0 \gamma _2\right)+\frac{1}{24} \gamma _0 L^4 \left(\beta _0+\gamma _0\right) \left(2 \beta _0+\gamma _0\right) \left(3 \beta _0+\gamma _0\right)\nonumber\\&\bs+L^3 \left(\frac{1}{6} \left(\beta _1 \gamma _0 \left(5 \beta _0+3 \gamma _0\right)+3 \gamma _1 \left(\beta _0+\gamma _0\right) \left(2 \beta _0+\gamma _0\right)\right)\right)\Bigg)\nonumber\\&+x^5\Bigg(\gamma _4 L+\frac{1}{2} L^2 \left(\beta _3 \gamma _0+3 \beta _1 \gamma _2+2 \gamma _1 \left(\beta _2+\gamma _2\right)+2 \gamma _3 \left(2 \beta _0+\gamma _0\right)\right)\nonumber\\&\bs+\frac{1}{6} L^3 \left(3 \beta _1^2 \gamma _0+\beta _1 \gamma _1 \left(14 \beta _0+9 \gamma _0\right)+3 \left(2 \beta _0+\gamma _0\right) \left(\beta _2 \gamma _0+\gamma _2 \left(2 \beta _0+\gamma _0\right)+\gamma _1^2\right)\right)\nonumber\\&\bs+\frac{1}{12} L^4 \left(\beta _1 \gamma _0 \left(13 \beta _0 \gamma _0+13 \beta _0^2+3 \gamma _0^2\right)+2 \gamma _1 \left(\beta _0+\gamma _0\right) \left(2 \beta _0+\gamma _0\right) \left(3 \beta _0+\gamma _0\right)\right)\nonumber\\&\bs+\frac{1}{120} \gamma _0 L^5 \left(\beta _0+\gamma _0\right) \left(2 \beta _0+\gamma _0\right) \left(3 \beta _0+\gamma _0\right) \left(4 \beta _0+\gamma _0\right)\Bigg)\Bigg\rbrace+\order{\alpha_s^6}\,,
			\label{eq:massmu}
		\end{align}
		where $x\equiv x(\mu^2)$, $L=\log(\mu^2/q^2)$, $\beta_i$ are the QCD beta function coefficients and $\gamma_i$ are the coefficients of the anomalous dimension of the quark mass.\par
		The QCD beta function coefficients are known to five-loop \cite{vanRitbergen:1997va,Gross:1973id,Caswell:1974gg, Jones:1974mm,Tarasov:1980au,Larin:1993tp,Czakon:2004bu,Baikov:2016tgj,Herzog:2017ohr} and their analytic expressions for the active flavor $n_f=3$ are:
		\begin{align}
			&\beta_0 =9/4\,,\quad\beta_1=4\,,\quad
			\beta_2 =3863/384\,,
			\beta_3 = 140599/4608 + (445 \zeta (3))/32\nonumber\\&
			\beta_{4} =\frac{11059213 \zeta (3)}{55296}-\frac{534385 \zeta (5)}{3072}-\frac{801 \pi ^4}{2048}+\frac{139857733}{1327104}\,.
		\end{align}
		The known five-loop quark mass anomalous dimension coefficients\cite{Tarrach:1980up,Tarasov:1982plg,Larin:1993tq,Vermaseren:1997fq,Chetyrkin:1997dh,Baikov:2014qja,Luthe:2016ima,Luthe:2016xec} for $n_f=3$ are:
		\begin{align}
			\gamma_m^{(0)}=&1 \,, \quad \gamma_m^{(1)}=\frac{91}{24}\,, \quad				\gamma_m^{(2)}=\frac{8885}{576}-\frac{5 \zeta (3)}{2}\,,\quad
			\gamma_m^{(3)}=-\frac{9295 \zeta (3)}{432}-\frac{125 \zeta (5)}{12}+\frac{3 \pi ^4}{32}+\frac{2977517}{41472}\,,\nonumber\\
			\gamma_m^{(4)} = &\frac{156509815}{497664}-\frac{23663747 \zeta (3)}{124416}+85 \zeta (3)^2+\frac{4753 \pi ^4}{4608}+\frac{118405 \zeta (7)}{576}-\frac{22625465 \zeta (5)}{62208}+\frac{125 \pi ^6}{2016}\,.
		\end{align}
		We also need the vacuum anomalous dimension for dimension-4 operators, which has been recently computed to five-loop\cite{Baikov:2018nzi}. Their analytic expression of the diagonal component relevant for this article is given by :
		\begin{align}
			\hat{\gamma}_0^{di}&\equiv  \frac{3}{16\pi^2}\gamma^{ii}_n x^n\nonumber\\&=\frac{3}{16\pi^2}\Bigg\lbrace-1-\frac{4 x}{3}+x^2 \left(\frac{2 \zeta (3)}{3}-\frac{223}{72}\right)+x^3 \left(\frac{346 \zeta (3)}{9}-\frac{1975 \zeta (5)}{54}+\frac{13 \pi ^4}{540}-\frac{3305}{1296}\right)\nn\\&\bs+x^4 \Bigg(\frac{6121 \zeta (3)^2}{864}-\frac{11881 \pi ^4}{8640}+\frac{1680599 \zeta (3)}{2592}+\frac{36001 \zeta (7)}{96}+\frac{93925 \pi ^6}{326592}-\frac{59711 \zeta (5)}{48}+\frac{16141627}{248832}\Bigg) \Bigg\rbrace\,.
		\end{align}
		\section{The RGSPT Coefficients Relevant The Dimension-0 and The Dimension-2 Adler Functions} \label{app:summed_sol}
		The first three summed series coefficients are presented as
		\begin{align}
			S_0[w]=& w^{-n_2 \tilde{\gamma }_0-n_1}\\S_1[w]=&w^{-n_2 \tilde{\gamma }_0-n_1-1} \Bigg(T_{1,0}-n_1 \tilde{\beta }_1 L_w+n_2 \times \big(-\tilde{\beta }_1 \tilde{\gamma }_0+\tilde{\gamma }_1+w \tilde{\beta }_1 \tilde{\gamma }_0-\tilde{\beta }_1 \tilde{\gamma }_0 L_w-w \tilde{\gamma }_1\big)\Bigg)\\
			S_2[w]=&w^{-n_2 \tilde{\gamma }_0-n_1-2}\Bigg\lbrace T_{2,0}+T_{1,0} \left(n_2 \left(\tilde{\gamma }_1-\tilde{\beta }_1 \tilde{\gamma }_0\right)+n_2 w \left(\tilde{\beta }_1 \tilde{\gamma }_0-\tilde{\gamma }_1\right)+L_w \left(-n_2 \tilde{\beta }_1 \tilde{\gamma }_0-n_1 \tilde{\beta }_1-\tilde{\beta }_1\right)\right)\nonumber\\&+\bigg\lbrace n_1 \left(\tilde{\beta }_2-\tilde{\beta }_1^2\right)+n_2 \left(-\frac{1}{2} \tilde{\beta }_1^2 \tilde{\gamma }_0-\frac{1}{2} \tilde{\beta }_1 \tilde{\gamma }_1+\frac{1}{2} \tilde{\beta }_2 \tilde{\gamma }_0+\frac{\tilde{\gamma }_2}{2}\right)+w \bigg[n_1 \left(\tilde{\beta }_1^2-\tilde{\beta }_2\right)+n_2 \left(\tilde{\beta }_1^2 \tilde{\gamma }_0-\tilde{\beta }_2 \tilde{\gamma }_0\right)\nonumber\\&+n_2^2 \left(-\tilde{\beta }_1^2 \tilde{\gamma }_0^2+2 \tilde{\beta }_1 \tilde{\gamma }_1 \tilde{\gamma }_0-\tilde{\gamma }_1^2\right)+L_w \big(n_2^2 \left(\tilde{\beta }_1 \tilde{\gamma }_0 \tilde{\gamma }_1-\tilde{\beta }_1^2 \tilde{\gamma }_0^2\right)+n_1 n_2 \left(\tilde{\beta }_1 \tilde{\gamma }_1-\tilde{\beta }_1^2 \tilde{\gamma }_0\right)\big)\bigg]\nonumber\\&+L_w^2 \bigg[\frac{1}{2} n_2^2 \tilde{\beta }_1^2 \tilde{\gamma }_0^2+\frac{1}{2} n_1^2 \tilde{\beta }_1^2+\frac{1}{2} n_2 \tilde{\beta }_1^2 \tilde{\gamma }_0+n_1 \left(n_2 \tilde{\beta }_1^2 \tilde{\gamma }_0+\frac{\tilde{\beta }_1^2}{2}\right)\bigg]+w^2 \bigg[n_2^2 \left(\frac{1}{2} \tilde{\beta }_1^2 \tilde{\gamma }_0^2-\tilde{\beta }_1 \tilde{\gamma }_1 \tilde{\gamma }_0+\frac{\tilde{\gamma }_1^2}{2}\right)\nonumber\\&+n_2 \left(-\frac{1}{2} \tilde{\beta }_1^2 \tilde{\gamma }_0+\frac{1}{2} \tilde{\beta }_1 \tilde{\gamma }_1+\frac{1}{2} \tilde{\beta }_2 \tilde{\gamma }_0-\frac{\tilde{\gamma }_2}{2}\right)\bigg]\bigg\rbrace\Bigg\rbrace
		\end{align}
		where $\tilde{X}_i\equiv X_i/\beta_0$, and the rest of them can be found by solving eq.~\eqref{summed_de} with the boundary conditions that $S_i[1]=T_{i,0}$ and for simplification of the expressions, we have taken $T_{0,0}=1$.
		\section{Perturbative coefficients Relevant for the Dimension-4 Corrections and their RGSPT coefficients}\label{app:dim4corrections}
		The RG-inaccessible coefficients needed for the dimension-4 operators are calculated in\ ref.~\cite{Chetyrkin:1985kn,Generalis:1989hf,Broadhurst:1985js,Generalis:1990iy,Loladze:1985qk,Becchi:1980vz,Surguladze:1990sp,Bagan:1985zp,Pascual:1981jr,Jamin:1992se,Generalis:1990id} and their values are:
		\begin{align}
			&p0_0=0,\quad p0_1=1,\quad p0_2=7/6,\quad q0_0=1,\quad q0_1=-1,\quad q0_2=-131/24,\quad hl0_0=1,\quad kl0_0=1\nonumber\\
			&t0_0=0,\quad t0_1=1,\quad t0_2=17/2,\quad h0_0=1,\quad g0_0=1,\quad g0_1=94/9-4/3\zeta_3,\quad k0_0=0,\quad k0_1=1\,.
		\end{align}
		\par Perturbative coefficients involving the condensates terms described in section~\ref{sec:dim_4_Adler} are:
		\begin{align}
			&p^{L+T}_0(w)=0,\quad p^{L+T}_1(w)=p0_1,\quad p^{L+T}_2(w)=\frac{\beta _1 p0_1}{\beta _0}+\frac{p0_2-\frac{\beta _1 p0_1}{\beta _0}}{w}\,,\nonumber\\&r^{L+T}_0(w)=0,\quad r^{L+T}_1(w)=\frac{\gamma _0 p0_1}{6 \beta _0 w}-\frac{\gamma _0 p0_1}{6 \beta _0},\quad r^{L+T}_2(w)=\frac{\frac{\beta _1 \gamma _0 p0_1}{6 \beta _0^2}-\frac{\gamma _0 p0_2}{6 \beta _0}}{w}\,,\nonumber\\&
			q^{L+T}_0(w)=q0_0,\quad q^{L+T}_1(w)=\frac{q0_1}{w},\quad q^{L+T}_2(w)=\frac{q0_2-\frac{\beta _1 q0_1 \log (w)}{\beta _0}}{w^2}\,,\nonumber\\&
			t^{L+T}_0(w)=0,\quad t^{L+T}_1(w)=\frac{t0_1}{w},\quad t^{L+T}_2(w)=\frac{t0_2-\frac{\beta _1 t0_1 \log (w)}{\beta _0}}{w^2}\,.
		\end{align}
		\par
		The RGSPT coefficients for the coefficients of $m^4$ to $\order{\alpha_s}$ are:
		\begin{align}
			k^{L+T}_0(w)=&\frac{\gamma^{ii}_0 t0_1 \left(2 (1-w) w^{-\frac{4\gamma _0}{\beta _0}}-2 w^{-\frac{4\gamma _0}{\beta _0}}+2\right)}{10 w \left(2 4\gamma _0-2 \beta _0\right)}\,,\quad g^{L+T}_0(w)=g0_0 w^{-\frac{4\gamma _0}{\beta _0}},\quad
			j^{L}_0(w)=0,\quad j^{L}_1(w)=0\,,\\
			k^{L+T}_1(w)=&\frac{\gamma^{ii}_0 w^{-\frac{4\gamma _0}{\beta _0}-2}}{10 \beta _0 \left(\beta _0-4\gamma _0\right){}^2} \Bigg(\beta _1 t0_1 \left(w^{\frac{4\gamma _0}{\beta _0}} \left(\beta _0 (\log (w)+1)-4\gamma _0 \log (w)\right)-\beta _0 w\right)-\beta _0 t0_2 \left(\beta _0-4\gamma _0\right) \left(w^{\frac{4\gamma _0}{\beta _0}}-w\right)\Bigg)\nonumber\\&+k0_1 w^{-\frac{4\gamma _0}{\beta _0}-1}+\frac{\frac{\gamma^{ii}_1 t0_1}{w}-\gamma^{ii}_1 t0_1 w^{-\frac{4\gamma _0}{\beta _0}-1}}{10 4\gamma _0}\,\\
			h^{L+T}_0(w)=&h0_0 w^{-\frac{4 \gamma _0}{\beta _0}}+\frac{q0_1 \left(\frac{1}{w}-w^{-\frac{4 \gamma _0}{\beta _0}}\right)}{2 \left(\beta _0-4 \gamma _0\right)}-\frac{q0_0 \gamma_1 w^{-\frac{4 \gamma _0}{\beta _0}} \left(\beta _0 \left(w^{\frac{4 \gamma _0}{\beta _0}}-1\right)-4 \gamma _0 (w-1)\right)}{2 \beta _0 \gamma _0 \left(\beta _0-4 \gamma _0\right)}\nonumber\\&\frac{q0_0 w^{-\frac{4 \gamma _0}{\beta _0}} \left(\beta _0^2 \left(-4 \beta _0+3 \beta _1+16 \gamma _0\right) \left(w^{\frac{4 \gamma _0}{\beta _0}}-1\right)+12 \beta _1 \gamma _0 \left(4 \gamma _0 (-w+\log (w)+1)-\beta _0 \log (w)\right)\right)}{24 \beta _0^2 \gamma _0 \left(\beta _0-4 \gamma _0\right)}	\,,\\
			h^{L+T}_1(w)=&w^{-\frac{4 \gamma _0}{\beta _0}-1} \left(\frac{h0_0 \left(4 \beta _0 \gamma _1-4 \beta _1 \gamma _0 (\log (w)+1)\right)}{\beta _0^2}+h0_1+\frac{\beta _1 q0_0 \log (w) \left(3 \beta _1 \gamma _0 (\log (w)+2)-2 \beta _0 \left(\beta _0+3 \gamma _1\right)\right)}{3 \beta _0^4}\right)\nonumber\\&+\frac{q0_0 \left(\beta _0^3 \left(-4 \beta _1+16 \gamma _1+6 \gamma _2\right)+2 \beta _0^2 \left(3 \left(\beta _2 \gamma _0+4 \gamma _1^2\right)-\beta _1 \left(8 \gamma _0+9 \gamma _1\right)\right)+6 \beta _1 \beta _0 \gamma _0 \left(\beta _1-8 \gamma _1\right)\right) w^{-\frac{4 \gamma _0}{\beta _0}-1}}{6 \beta _0^4\left(\beta _0+4 \gamma _0\right)}\nonumber\\&+\frac{q0_0 \left(3 \beta _2 \gamma _0+\gamma _1 \left(4 \beta _0-3 \beta _1-16 \gamma _0+12 \gamma _1\right)-12 \gamma _2 \gamma _0\right)}{6 \gamma _0 \left(\beta _0^2-16 \gamma _0^2\right)}+\frac{\gamma^{ii}_2 q0_0}{2\left(\beta _0-4 \gamma _0\right)}-\frac{ q0_0 \left(\gamma^{ii}_2-8 \frac{\beta _1^2 \gamma _0^2}{\beta_0^4}\right) w^{-\frac{4 \gamma _0}{\beta _0}-1}}{2\left(\beta _0+4 \gamma _0\right)}\nonumber\\
			&+\frac{1}{\left(\beta _0-4 \gamma _0\right)}\bigg\lbrace\frac{q0_0 \left(\beta _0^3 \gamma _2-\beta _0^2 \left(\beta _2 \gamma _0+\gamma _1 \left(\beta _1+4 \gamma _1\right)\right)+\beta _1 \beta _0 \gamma _0 \left(\beta _1+8 \gamma _1\right)-4 \beta _1^2 \gamma _0^2\right) w^{1-\frac{4 \gamma _0}{\beta _0}}}{\beta _0^4}\nonumber\\&+\frac{\frac{q0_1 \left(-4 \beta _0+3 \beta _1+16 \gamma _0-12 \gamma _1\right)}{24 \gamma _0}+w^{-\frac{4 \gamma _0}{\beta _0}} \left(q0_1 \left(\frac{\beta _0}{6 \gamma _0}-\frac{\beta _1-4 \gamma _1}{8 \gamma _0}-\frac{2 \gamma _1}{\beta _0}+\frac{2 \beta _1 \gamma _0 (\log (w)+1)}{\beta _0^2}-\frac{2}{3}\right)-\frac{q0_2}{2}\right)}{w}\nonumber\\&+\frac{\frac{q0_2}{2}-\frac{\beta _1 q0_1 \log (w)}{2 \beta _0}}{w^2}+\frac{2 q0_1 \left(\beta _0 \gamma _1-\beta _1 \gamma _0\right) w^{-\frac{4 \gamma _0}{\beta _0}}}{\beta _0^2}\bigg\rbrace\\
			g^{L+T}_1(w)&=-\frac{\beta _1 4\gamma _0 g0_0 \log (w) w^{-\frac{4\gamma _0}{\beta _0}-1}}{\beta _0^2}+\frac{w^{-\frac{4\gamma _0}{\beta _0}-1} \left(4 \beta _0^2 g0_1+4 \beta _0 \gamma _1 g0_0 (1-w)-4 \beta _1 4\gamma _0 g0_0 (1-w)\right)}{4 \beta _0^2}\,\\
			h^{L}_0(w)=&-\frac{\gamma^{ii}_0 w^{-\frac{4\gamma _0}{\beta _0}} \left(\beta _0^2 \left(\beta _1-\gamma _1\right) \left(w^{\frac{4\gamma _0}{\beta _0}}-1\right)-\beta _0 4\gamma _0 \left(\beta _1 \log (w)-\gamma _1 (w-1)\right)+\beta _1 4\gamma _0^2 (-w+\log (w)+1)\right)}{8 \beta _0^2 4\gamma _0 \left(\beta _0-4\gamma _0\right)}\nonumber\\&+hl0_0 w^{-\frac{4\gamma _0}{\beta _0}}+\frac{\gamma^{ii}_1 \left(1-w^{-\frac{4\gamma _0}{\beta _0}}\right)}{8 4\gamma _0}-\frac{\gamma^{ii}_0(1- w^{1-\frac{4\gamma _0}{\beta _0}})}{8 x \left(\beta _0-4\gamma _0\right)}\,,
		\end{align}
		\begin{align}
			h^{L}_1(w)=&w^{-\frac{4 \gamma _0}{\beta _0}} \bigg(-\frac{2 \beta _1^2 \gamma _0}{\beta _0^4}+\frac{4 \beta _1 \gamma _1}{\beta _0^3}+\frac{\beta _1 \gamma _1}{2 \beta _0^2 \gamma _0}-\frac{2 \gamma _1}{3 \beta _0 \gamma _0}-\frac{2 \gamma _1^2}{\beta _0^2 \gamma _0}+\frac{2 \beta _1}{3 \beta _0^2}-\frac{\beta _2}{2 \beta _0^2}+hl0_0 \left(\frac{4 \beta _1 \gamma _0}{\beta _0^2}-\frac{4 \gamma _1}{\beta _0}\right)+\frac{\frac{2 \gamma _1}{3 \gamma _0}+\frac{\gamma^{ii}_2}{2}}{\left(\beta _0+4 \gamma _0\right)}\nonumber\\&+\left(\frac{2 \beta _1 \gamma _1}{\beta _0^3}-\frac{2 \beta _1^2 \gamma _0}{\beta _0^4}\right) \log (w)\bigg)+\frac{\left(\beta _0^3 \gamma _2-\beta _0^2 \left(\beta _2 \gamma _0+\gamma _1 \left(\beta _1+4 \gamma _1\right)\right)+\beta _1 \beta _0 \gamma _0 \left(\beta _1+8 \gamma _1\right)-4 \beta _1^2 \gamma _0^2\right) w^{1-\frac{4 \gamma _0}{\beta _0}}}{\beta _0^4 \left(\beta _0-4 \gamma _0\right)}\nonumber\\&+\frac{-\frac{\gamma _1 \left(\beta _1-4 \gamma _1\right)}{\gamma _0}+\beta _2-4 \gamma _2}{2 \left(\beta _0^2-16 \gamma _0^2\right)}+w^{-\frac{4 \gamma _0}{\beta _0}-1} \Bigg\lbrace hl0_0 \left(-\frac{4 \beta _1 \gamma _0}{\beta _0^2}+\frac{4 \gamma _1}{\beta _0}-\frac{4 \beta _1 \gamma _0 \log (w)}{\beta _0^2}\right)+hl0_1+\frac{\beta _1^2 \gamma _0 \log ^2(w)}{\beta _0^4}\nonumber\\&+\left(\frac{2 \beta _1^2 \gamma _0}{\beta _0^4}-\frac{2 \beta _1 \gamma _1}{\beta _0^3}-\frac{2 \beta _1}{3 \beta _0^2}\right) \log (w)+\frac{1}{6 \beta _0^4 \left(\beta _0+4 \gamma _0\right)}\bigg\lbrace2 \beta _0^2 \left(3 \left(\beta _2 \gamma _0+4 \gamma _1^2\right)-\beta _1 \left(8 \gamma _0+9 \gamma _1\right)\right)\nonumber\\&\bs+\beta _0^3 \left(-4 \beta _1+16 \gamma _1+6 \gamma _2\right)+6 \beta _1 \beta _0 \gamma _0 \left(\beta _1-8 \gamma _1\right)+24 \beta _1^2 \gamma _0^2-3 \beta _0^4 \gamma^{ii}_2\bigg\rbrace\Bigg\rbrace\,,\\
			k^{L}_0(w)=&-\frac{w^{1-\frac{4 \gamma _0}{\beta _0}}-1}{3 \left(\beta _0-4 \gamma _0\right) x}-\frac{1}{9 \gamma _0}+\frac{\beta _1-4 \gamma _1}{12 \gamma _0 \left(\beta _0-4 \gamma _0\right)}+\frac{4 \left(\beta _0 \gamma _1-\beta _1 \gamma _0\right) w^{1-\frac{4 \gamma _0}{\beta _0}}}{3 \beta _0^2 \left(\beta _0-4 \gamma _0\right)}\nonumber\\&\bs\bs+w^{-\frac{4 \gamma _0}{\beta _0}}\left(kl0_0+\frac{-3 \beta _0 \left(\beta _1-4 \gamma _1\right)+4 \beta _0^2-12 \beta _1 \gamma _0 (\log (w)+1)}{36 \beta _0^2 \gamma _0}\right)\,,\\
			k^{L}_1(w)=&kl0_1 w^{-\frac{4 \gamma _0}{\beta _0}-1}+\frac{2 \beta _1 \log (w) w^{-\frac{4 \gamma _0}{\beta _0}-1} \left(2 \beta _0 \gamma _1 w+\beta _1 \gamma _0 (\log (w)-2 w)\right)}{3 \beta _0^4}+\frac{\gamma _1 \left(4 \gamma _1-\beta _1\right)+\gamma _0 \left(\beta _2-4 \gamma _2\right)}{3 \gamma _0 \left(\beta _0^2-16 \gamma _0^2\right)}\nonumber\\&+\frac{3 \gamma _0 \gamma^{ii}_2+4 \gamma _1}{9 \gamma _0 \left(\beta _0+4 \gamma _0\right)}-\frac{4 \beta _1 \log (w) \left(3 \beta _0 \gamma _1-3 \beta _1 \gamma _0+\beta _0^2 \left(9 \gamma _0 kl0_0+1\right)\right) w^{-\frac{4 \gamma _0}{\beta _0}-1}}{9 \beta _0^4}+\frac{w^{-\frac{4 \gamma _0}{\beta _0}-1}}{9 \beta _0^4\left(\beta _0+4 \gamma _0\right)} \bigg\lbrace6 \beta _1 \beta _0 \gamma _0 \nonumber\\&\times\left(\beta _1-8 \gamma _1\right)-2 \beta _0^2 \left(\beta _1 \left(9 \gamma _1+8 \gamma _0 \left(9 \gamma _0 kl0_0+1\right)\right)-3 \left(\beta _2 \gamma _0+4 \gamma _1^2\right)\right)+\beta _0^4 \left(36 \gamma _1 kl0_0-3 \gamma^{ii}_2\right)+24 \beta _1^2 \gamma _0^2\nonumber\\&+2 \beta _0^3 \left(3 \gamma _2-2 \left(\beta _1-4 \gamma _1\right) \left(9 \gamma _0 kl0_0+1\right)\right)\bigg\rbrace-\frac{4 w^{-\frac{4 \gamma _0}{\beta _0}-1}}{9 \beta _0^4}\beta _1 \log (w) \left(3 \beta _0 \gamma _1-3 \beta _1 \gamma _0+\beta _0^2 \left(9 \gamma _0 kl0_0+1\right)\right)\nonumber\\&+\frac{2 w^{1-\frac{4 \gamma _0}{\beta _0}}}{3 \beta _0^4 \left(\beta _0-4 \gamma _0\right)}\left(\beta _0^3 \gamma _2-\beta _0^2 \left(\beta _2 \gamma _0+\gamma _1 \left(\beta _1+4 \gamma _1\right)\right)+\beta _1 \beta _0 \gamma _0 \left(\beta _1+8 \gamma _1\right)-4 \beta _1^2 \gamma _0^2\right)+\frac{w^{-\frac{4 \gamma _0}{\beta _0}}}{9 \beta _0^4 \gamma _0}\bigg\lbrace24 \beta _1 \beta _0 \gamma _0 \gamma _1\nonumber\\&-12 \beta _1^2 \gamma _0^2-4 \beta _0^3 \gamma _1 \left(9 \gamma _0 kl0_0+1\right)+\beta _0^2 \left(\beta _1 \left(3 \gamma _1+4 \gamma _0 \left(9 \gamma _0 kl0_0+1\right)\right)-3 \left(\beta _2 \gamma _0+4 \gamma _1^2\right)\right)\bigg\rbrace\\\nonumber\\
			\text{where }w&=(1-\beta_0 x L)\nonumber\,.
		\end{align}

		\section{Contributions to the Adler Function }\label{app:adlercoef}
		\subsection{Dimension-zero contributions}
		In massless case, Adler function is known to $\mathcal{O}(\alpha^4_s)$\cite{Appelquist:1973uz,Zee:1973sr,Chetyrkin:1979bj,Dine:1979qh,Gorishnii:1990vf,Surguladze:1990tg,Chetyrkin:1996ez,Baikov:2008jh,Baikov:2010je,Herzog:2017dtz} and contribution from longitudinal part is zero ($D_{i,j}^{L,(0)}=0$) while $D^{L+T,0}_{i,j}$ is given by:
		\begin{align}
			\mathcal{D}^{L+T,0}_{i,j}=&1+x+x^2 \left(\frac{299}{24}-9 \zeta (3)\right)+x^3 \left(-\frac{779 \zeta (3)}{4}+\frac{75 \zeta (5)}{2}+\frac{58057}{288}\right)\nonumber\\&+x^4 \left(\frac{4185 \zeta (3)^2}{8}+\frac{729 \pi ^2 \zeta (3)}{16}-\frac{1704247 \zeta (3)}{432}+\frac{34165 \zeta (5)}{96}-\frac{1995 \zeta (7)}{16}-\frac{13365 \pi ^2}{256}+\frac{78631453}{20736}\right)
		\end{align}
		
		\subsection{The Dimension-2 Corrections}\label{app:dim2adler}
		The dimension-2 correction to the Adler function for the quark flavor $i$ and $j$ is known to $\mathcal{O}\left(\alpha^3_s\right)$\cite{Baikov:2004ku,Baikov:2004tk,Chetyrkin:1993hi,Gorishnii:1986pz,Generalis:1989hf,Bernreuther:1981sp} and the analytic expression reads:
		\begin{align}
			\mathcal{D}_{2,ij}^{L+T,V/A}(s)&=\frac{3}{4 \pi ^2 s} \Bigg\lbrace \left(m_u^2+m_d^2+m_s^2\right)\big(x^2 \left(\frac{8 \zeta (3)}{3}-\frac{32}{9}\right)+x^3 \left(4 \zeta (3)^2+\frac{1592 \zeta (3)}{27}-\frac{80 \zeta (5)}{27}-\frac{2222}{27}\right)\big)\nonumber\\&\bs\bs+\left(m_i^2+m_j^2\right)\bigg(1+\frac{13 x}{3}+x^2 \left(\frac{179 \zeta (3)}{54}-\frac{520 \zeta (5)}{27}+\frac{23077}{432}\right)+x^3 \bigg(\frac{53 \zeta (3)^2}{2}\nonumber\\&\bs\bs\bs\bs\bs-\frac{1541 \zeta (3)}{648}+\frac{79835 \zeta (7)}{648}-\frac{54265 \zeta (5)}{108}-\frac{\pi ^4}{36}+\frac{3909929}{5184}\bigg)\bigg) \nonumber\\&\bs\bs\pm  m_i m_j \Bigg(\frac{2 x}{3}+x^2 \left(-\frac{55 \zeta (3)}{27}-\frac{5 \zeta (5)}{27}+\frac{769}{54}\right)+x^3 \Big(-\frac{11677 \zeta (3)^2}{108}\nonumber\\&\bs\bs\bs\bs\bs+\frac{70427 \zeta (3)}{324}+\frac{82765 \zeta (5)}{54}-\frac{555233 \zeta (7)}{864}+\frac{\pi^4}{9}-\frac{7429573}{3888}\Big)\Bigg)\Bigg\rbrace\,,
		\end{align}
		where upper and lower signs correspond to vector and axial-vector components, respectively, and this convention is used for the Alder functions in this section. It should be noted that the $\mathcal{O}(\alpha^4_s)$ correction to $m_i m_j$ term has been obtained from eq(15) of ref.~\cite{Baikov:2004tk}.
		The longitudinal component of the dimension-2 operator is known to $\order{\alpha_s^4}$\cite{Becchi:1980vz,Broadhurst:1981jk,Chetyrkin:1996sr,Baikov:2005rw,Gorishnii:1990zu,Gorishnii:1991zr} and has the form:
		\begin{align}
			\mathcal{D}_{2,ij}^{L,V/A}=&\frac{-3}{8\pi^2}\frac{\left(m_i\mp m_j\right)^2}{M^2_{\tau}}\Bigg\lbrace1+\frac{17 x}{3}+x^2 \left(\frac{9631}{144}-\frac{35 \zeta (3)}{2}\right)+x^3 \Big(-\frac{91519 \zeta (3)}{216}+\frac{715 \zeta (5)}{12}-\frac{\pi ^4}{36}+\frac{4748953}{5184}\Big)\nonumber\\&\bs\bs\bs\bs+x^4 \Big(\frac{192155 \zeta (3)^2}{216}-\frac{46217501 \zeta (3)}{5184}+\frac{455725 \zeta (5)}{432}-\frac{125 \pi ^6}{9072}-\frac{52255 \zeta (7)}{256}\nonumber\\&\bs\bs\bs\bs\bs-\frac{3491 \pi ^4}{10368}+\frac{7055935615}{497664}\Big)\Bigg\rbrace\,.
		\end{align}
		From the above dimension-2 Alder functions, the important piece relevant for Cabibbo suppressed strange quark mass determination\cite{Pich:1999hc} using eq.~\eqref{eq:delRkl} are:
		\begin{align}
			\delta \mathcal{D}^{L+T,V+A}_{2}(s)&=(\mathcal{D}^{L+T,V+A}_{2,ud}(s)-\mathcal{D}^{L+T,V+A}_{2,us}(s))\nonumber\\&=\frac{-3 m^2_s}{2\pi^2 s}(1-\epsilon^2_d) \Bigg(1+\frac{13 x}{3}+x^2 \left(\frac{179 \zeta (3)}{54}-\frac{520 \zeta (5)}{27}+\frac{23077}{432}\right)\nonumber\\&\bs\bs\bs\hs\hs+x^3 \left(\frac{53 \zeta (3)^2}{2}-\frac{1541 \zeta (3)}{648}+\frac{79835 \zeta (7)}{648}-\frac{54265 \zeta (5)}{108}-\frac{\pi ^4}{36}+\frac{3909929}{5184}\right)\Bigg)\\&\equiv\frac{-3 m^2_s(-\xi^2s)}{2\pi^2 s}(1-\epsilon^2_d)\sum_{i=0}\tilde{d}^{L+T}_{i,0}(\xi^2)x(-\xi^2s)^i
			\label{eq:su3adlerLT}
			\\&=\frac{-3 m^2_s(\xi^2\mtsq)}{2\pi^2 s}(1-\epsilon^2_d)\sum_{i=0}^{4}\sum_{j=0}^{i}\tilde{d}^{L+T}_{i,j}x(\xi^2\mtsq)^i \log^j\left(\frac{\xi^2\mtsq}{-s}\right)\,,
			\label{eq:su3adlerLT1}
		\end{align}
		and the corresponding contribution from the longitudinal component is:
		\begin{align}
			\delta \mathcal{D}^{L,V+A}_{2}(s)&=\mathcal{D}^{L,V+A}_{2,ud}-\mathcal{D}^{L,V+A}_{2,us}\nonumber\\&=\frac{3 m^2_s}{4\pi^2 M^2_{\tau}}(1-\epsilon^2_d) \Biggl\lbrace1+\frac{17 x}{3}+x^2 \left(\frac{9631}{144}-\frac{35 \zeta (3)}{2}\right)+x^3 \Big(-\frac{91519 \zeta (3)}{216}+\frac{715 \zeta (5)}{12}-\frac{\pi ^4}{36}\nonumber\\&\bs\bs\bs+\frac{4748953}{5184}\Big)+x^4 \Big(\frac{192155 \zeta (3)^2}{216}-\frac{46217501 \zeta (3)}{5184}+\frac{455725 \zeta (5)}{432}-\frac{52255 \zeta (7)}{256}\nonumber\\&\bs\bs\bs-\frac{125 \pi ^6}{9072}-\frac{3491 \pi ^4}{10368}+\frac{7055935615}{497664}\Bigg)\Biggr\rbrace\\&\equiv\frac{3 m^2_s(-\xi^2s)}{4\pi^2 M^2_{\tau}}(1-\epsilon^2_d)\sum_{i=0}\tilde{d}^{L}_{i,0}(\xi^2)x(-\xi^2s)^i\label{eq:su3adlerL}\\&=\frac{3 m^2_s(\xi^2\mtsq)}{4\pi^2 M^2_{\tau}}(1-\epsilon^2_d)\sum_{i=0}^{4}\sum_{j=0}^{i}\tilde{d}^{L}_{i,j}x(\xi^2\mtsq)^i \log^j\left(\frac{\xi^2\mtsq}{-s}\right)\,.
			\label{eq:su3adlerL1}
		\end{align}
		The RGSPT coefficients for the dimension-2 operators can be written in the following form:
		\begin{align}
			\delta \mathcal{D}^{J,2}_{V+A}(s)&=\text{norm}\times\frac{ 3 m^2_s}{2\pi^2 }(1-\epsilon^2_d)\Bigg\lbrace\frac{1}{w^{8/9}}+x \left(\frac{\tilde{d}^{J}_{1,0}}{w^{17/9}}-\frac{1.79012}{w^{8/9}}+\frac{1.79012}{w^{17/9}}-\frac{1.58025 \log (w)}{w^{17/9}}\right)\nonumber\\&\bs+x^2 \Big(\frac{1.79012 \tilde{d}^{J}_{1,0}}{w^{26/9}}+\frac{\tilde{d}^{J}_{2,0}}{w^{26/9}}-\frac{1.79012 \tilde{d}^{J}_{1,0}}{w^{17/9}}+\frac{\left(-3.35802 \tilde{d}^{J}_{1,0}-8.82061\right) \log (w)}{w^{26/9}}-\frac{0.339459}{w^{8/9}}\nonumber\\&\bs\bs-\frac{4.36949}{w^{17/9}}+\frac{4.70895}{w^{26/9}}+\frac{2.65325 \log ^2(w)}{w^{26/9}}+\frac{2.82884 \log (w)}{w^{17/9}}\Big)\nonumber\\&\bs+x^3 \Big(\frac{6.01952 \tilde{d}^{J}_{1,0}}{w^{35/9}}+\frac{1.79012 \tilde{d}^{J}_{2,0}}{w^{35/9}}+\frac{\tilde{d}^{J}_{3,0}}{w^{35/9}}-\frac{0.339459 \tilde{d}^{J}_{1,0}}{w^{17/9}}-\frac{5.68006 \tilde{d}^{J}_{1,0}}{w^{26/9}}-\frac{1.79012 \tilde{d}^{J}_{2,0}}{w^{26/9}}\nonumber\\&\bs\bs+\frac{\left(8.62308 \tilde{d}^{J}_{1,0}+27.3673\right) \log ^2(w)}{w^{35/9}}+\frac{\left(6.01128 \tilde{d}^{J}_{1,0}+19.7019\right) \log (w)}{w^{26/9}}+\frac{0.593473}{w^{8/9}}\nonumber\\&\bs\bs-\frac{3.28306}{w^{17/9}}+\frac{\left(-15.1635 \tilde{d}^{J}_{1,0}-5.1358 \tilde{d}^{J}_{2,0}-39.8653\right) \log (w)}{w^{35/9}}-\frac{14.9321}{w^{26/9}}+\frac{17.6217}{w^{35/9}}\nonumber\\&\bs\bs-\frac{4.5422 \log ^3(w)}{w^{35/9}}-\frac{4.74965 \log ^2(w)}{w^{26/9}}+\frac{0.53643 \log (w)}{w^{17/9}}\Big)
			\nonumber\\&\bs+x^4 \Bigg(\frac{0.593473 \tilde{d}^{J}_{1,0}}{w^{17/9}}+\frac{27.6536 \tilde{d}^{J}_{1,0}}{w^{44/9}}+\frac{7.3301 \tilde{d}^{J}_{2,0}}{w^{44/9}}+\frac{1.79012 \tilde{d}^{J}_{3,0}}{w^{44/9}}+\frac{\tilde{d}^{J}_{4,0}}{w^{44/9}}-\frac{6.29286 \tilde{d}^{J}_{1,0}}{w^{26/9}}\nonumber\\&\bs\bs-\frac{0.339459 \tilde{d}^{J}_{2,0}}{w^{26/9}}-\frac{21.9542 \tilde{d}^{J}_{1,0}}{w^{35/9}}-\frac{6.99064 \tilde{d}^{J}_{2,0}}{w^{35/9}}-\frac{1.79012 \tilde{d}^{J}_{3,0}}{w^{35/9}}-\frac{12.673}{w^{8/9}}+\frac{11.6487}{w^{17/9}}\nonumber\\&\bs\bs+\frac{\left(1.13991 \tilde{d}^{J}_{1,0}+11.9782\right) \log (w)}{w^{26/9}}+\frac{\left(-19.8721 \tilde{d}^{J}_{1,0}-71.1438\right) \log ^3(w)}{w^{44/9}}-\frac{15.09}{w^{26/9}}\nonumber\\&\bs\bs+\frac{\left(-15.4364 \tilde{d}^{J}_{1,0}-59.0364\right) \log ^2(w)}{w^{35/9}}-\frac{60.9336}{w^{35/9}}+\frac{77.0479}{w^{44/9}}+\frac{8.13109 \log ^3(w)}{w^{35/9}}\nonumber\\&\bs\bs-\frac{\left(68.5739 \tilde{d}^{J}_{1,0}+21.5065 \tilde{d}^{J}_{2,0}+6.91358 \tilde{d}^{J}_{3,0}+192.701\right) \log (w)}{w^{44/9}}-\frac{0.937834 \log (w)}{w^{17/9}}\nonumber\\&\bs\bs+\frac{\left(39.8584 \tilde{d}^{J}_{1,0}+9.19372 \tilde{d}^{J}_{2,0}+111.714\right) \log (w)}{w^{35/9}}+\frac{7.85071 \log ^4(w)}{w^{44/9}}-\frac{0.900672 \log ^2(w)}{w^{26/9}}\nonumber\\&\bs\bs+\frac{\left(67.7471 \tilde{d}^{J}_{1,0}+17.7534 \tilde{d}^{J}_{2,0}+186.459\right) \log ^2(w)}{w^{44/9}}\Bigg)\Bigg\rbrace\,.
			\label{eq:summed2_1}
		\end{align}
		A more compact form for the above equation is:
		\begin{align}
			\delta \mathcal{D}^{J,2}_{V+A}(s)	\equiv\text{norm}\times\frac{ 3 m^2_s}{2\pi^2 }(1-\epsilon^2_d)\sum _{i=0}^4 \sum _{k=0}^i \sum _{j=0}^k x^i \hs\tilde{T}^{J}_{i,j,k} \frac{\log^j(w)}{w^{2\gamma_0/\beta_0+k}}\,,
			\label{eq:summed_D2}
		\end{align}
		where
		\begin{align}
			\text{norm}=
			\begin{cases}
				\frac{-1}{s},\bs \text{if } &J= 0+1\\
				\frac{1}{2\mtsq},\bs &J=0\,,
			\end{cases}
		\end{align}
		and $\tilde{d}_i^{J}$ can be obtained from eq.~\eqref{eq:su3adlerLT},\eqref{eq:su3adlerL}.
		\section{Details of \texorpdfstring{$m_s$}{} and \texorpdfstring{$\vert V_{us}\vert$}{} determinations in different schemes and the details of sources of uncertainty} \label{app:determination_details}
		\subsection{Strange quark mass  determinations using pQCD inputs}\label{app:pQCD_mass}
		In this section, the strange quark mass is calculated using the longitudinal component calculated using OPE as described in sec.~\ref{sec:dim_2_behaviour}. As mentioned before, these contributions are poorly convergent, and the strange quark mass determinations will suffer from the large truncation uncertainties in addition to significant dependence on the moments used. The dependence on the moment can be slightly reduced by using the prescription I at the cost of enhanced truncation uncertainty. This behavior in the different perturbative schemes and the details of various sources of uncertainties are discussed in the later subsections.
		\subsubsection{\bf{Strange quark mass determination using CIPT scheme}}
		Determination of $m_s(\mtsq)$ using CIPT are based on dimension-2 contributions described in section[\ref{sec:CIPT_intro}]. Using prescription-I, can see that dimension-2 contributions are truncated at $\order{\alpha_s^3},\order{\alpha_s^3},\order{\alpha_s^2}$ and $\order{\alpha_s}$ for $k=0,1,2,3$ and 4 respectively. This truncation result into the enhancement in the total uncertainty in the $m_s(\mtsq)$ determination and can be seen in the tables[\ref{tab:msCI},\ref{tab:msCI1}]. However, the main advantage of using prescription-I is that the masses from various moments, using different experimental inputs, agree within the uncertainty, which is not the case in using prescription-2. For this reason, we are presenting $m_s(\mtsq)$ from both prescriptions for different schemes in other schemes too.
		\begin{center}
			\begin{table}[H]
				\centering
				\begin{tabular}{|c|c|c|c|c|c|c|c|c|c|c|c|}
					\hline
					Parameter&\multicolumn{5}{c|}{Moments ALEPH\cite{Chen:2001qf}}&\multicolumn{5}{c|}{Moment OPAL\cite{OPAL:2004icu}} \\ \cline{2-6}\cline{7-11}
					\text{} &(0,0)&(1,0)&(2,0)&(3,0)&(4,0)&(0,0)&(1,0)&(2,0)&(3,0)&(4,0)\\ \hline
					$m_s(\mtsq)$&$135_{-37}^{+34}$&$122_{-24}^{+28}$&$120_{-24}^{+32}$&$105_{-23}^{+31}$&$108_{-24}^{+38}$&$124_{-31}^{+30}$&$106_{-26}^{+27}$&$105_{-25}^{+30}$&$94_{-22}^{+29}$&$100_{-24}^{+36}$\\\hline
					$\delta R^{kl}_{\tau}(\text{Exp.})$ &$\text{}^{+27.4}_{-34.8}$&$\text{}^{+15.4}_{-17.7}$&$\text{}^{+11.7}_{-13.0}$&$\text{}^{+9.4}_{-10.4}$&$\text{}^{+9.0}_{-9.9}$&$\text{}^{+23.2}_{-28.7}$&$\text{}^{+17.8}_{-21.6}$&$\text{}^{+14.4}_{-16.8}$&$\text{}^{+11.2}_{-12.8}$&$\text{}^{+10.6}_{-12.0}$\\\hline
					$\xi\in\left[.75,2.0\right]$&$\text{}^{+14.3}_{-6.2}$&$\text{}^{+17.3}_{-8.0}$&$\text{}^{+20.8}_{-10.4}$&$\text{}^{+21.1}_{-10.9}$&$\text{}^{+23.8}_{-13.4}$&$\text{}^{+13.4}_{-5.8}$&$\text{}^{+15.1}_{-7.1}$&$\text{}^{+18.5}_{-9.2}$&$\text{}^{+19.0}_{-9.9}$&$\text{}^{+22.1}_{-12.5}$\\\hline
					Truncation uncertainty&$\text{}^{-8.7}_{+10.7}$&$\text{}^{-9.4}_{+12.2}$&$\text{}^{-12.1}_{+17.3}$&$\text{}^{-11.4}_{+16.8}$&$\text{}^{-15.6}_{+26.7}$&$\text{}^{-8.0}_{+10.0}$&$\text{}^{-8.2}_{+10.6}$&$\text{}^{-10.7}_{+15.3}$&$\text{}^{-10.3}_{+15.2}$&$\text{}^{-14.5}_{+25.1}$\\\hline
					$s_0\in\left[3,\mtsq\right](\GeV^2)$&8.3&11.1&12.7&12.5&6.4&7.7&9.6&11.2&11.2&5.9\\\hline
				\end{tabular}
				\caption{Strange quark mass using CIPT in prescription I. Other sources of uncertainties are not shown in the table, but are added in the quadrature for $m_s(\mtsq)$ in the second row.}
				\label{tab:msCI}
			\end{table}
			\begin{table}[H]
				\centering
				\begin{tabular}{|c|c|c|c|c|c|c|c|c|c|c|c|}
					\hline
					Parameter&\multicolumn{5}{c|}{Moments ALEPH\cite{Chen:2001qf}}&\multicolumn{5}{c|}{Moment OPAL\cite{OPAL:2004icu}} \\ \cline{2-6}\cline{7-11}
					\text{}&(0,0)&(1,0)&(2,0)&(3,0)&(4,0)&(0,0)&(1,0)&(2,0)&(3,0)&(4,0)\\ \hline
					$m_s(\mtsq)$&$126_{-35}^{+31}$&$112_{-22}^{+26}$&$96_{-19}^{+25}$&$82_{-18}^{+25}$&$69_{-17}^{+24}$&$116_{-29}^{+27}$&$97_{-24}^{+25}$&$84_{-20}^{+24}$&$73_{-18}^{+23}$&$64_{-16}^{+23}$\\\hline
					$\delta R^{kl}_{\tau}(\text{Exp.})$ &$\text{}^{+25.7}_{-32.6}$&$\text{}^{+14.1}_{-16.3}$&$\text{}^{+9.5}_{-10.5}$&$\text{}^{+7.4}_{-8.1}$&$\text{}^{+5.9}_{-6.4}$&$\text{}^{+21.7}_{-26.9}$&$\text{}^{+16.4}_{-19.8}$&$\text{}^{+11.6}_{-13.6}$&$\text{}^{+8.8}_{-10.0}$&$\text{}^{+6.9}_{-7.8}$\\\hline
					$\xi\in\left[.75,2.0\right]$&$\text{}^{+13.2}_{-6.4}$&$\text{}^{+16.3}_{-8.3}$&$\text{}^{+18.0}_{-9.4}$&$\text{}^{+18.7}_{-10.0}$&$\text{}^{+18.6}_{-10.1}$&$\text{}^{+12.3}_{-6.0}$&$\text{}^{+14.3}_{-7.2}$&$\text{}^{+15.9}_{-8.3}$&$\text{}^{+16.8}_{-9.0}$&$\text{}^{+17.1}_{-9.3}$\\\hline
					Truncation uncertainty&$\text{}^{-7.3}_{+8.8}$&$\text{}^{-8.0}_{+10.1}$&$\text{}^{-8.0}_{+10.6}$&$\text{}^{-7.7}_{+10.6}$&$\text{}^{-7.2}_{+10.4}$&$\text{}^{-6.7}_{+8.1}$&$\text{}^{-6.9}_{+8.8}$&$\text{}^{-7.0}_{+9.3}$&$\text{}^{-6.9}_{+9.6}$&$\text{}^{-6.6}_{+9.6}$\\\hline
					$s_0\in\left[3,\mtsq\right](\GeV^2)$&7.8&10.2&10.2&9.7&9.1&7.2&8.8&9.0&8.7&8.5\\\hline
				\end{tabular}
				\caption{ CIPT determination of strange quark mass using in prescription II. Only significant sources of uncertainty are shown in the table separately, while the rest are added in the quadrature and appear in $m_s(\mtsq)$.}
				\label{tab:msCI1}
			\end{table}
		\end{center}

		\subsubsection{\bf{Strange quark mass determination using FOPT scheme}}
		The effect of different prescription used is very significant in the FOPT where perturbative contributions from dimension-2 Alder function are truncated at $\order{\alpha_s^4}$ and $\order{\alpha_s^2}$ for for the moment $k=0\text{ and }1$ while rest of them are truncated at $\order{\alpha_s}$. The final results for $m_s(\mtsq)$ determination using FOPT in prescriptions I and II are presented in tables~\ref{tab:msFOPT} and \ref{tab:msFOPT2}, respectively.
		\begin{center}
			\begin{table}[H]
				\centering
				\begin{tabular}{|c|c|c|c|c|c|c|c|c|c|c|c|}
					\hline
					Parameter&\multicolumn{5}{c|}{Moments ALEPH\cite{Chen:2001qf}}&\multicolumn{5}{c|}{Moment OPAL\cite{OPAL:2004icu}} \\ \cline{2-6}\cline{7-11}
					\text{} &(0,0)&(1,0)&(2,0)&(3,0)&(4,0)&(0,0)&(1,0)&(2,0)&(3,0)&(4,0)\\ \hline
					$m_s(\mtsq)$&$114_{-34}^{+33}$&$134^{+38}_{-30}$&$147^{+48}_{-34}$&$137^{+45}_{-31}$&$127_{-29}^{+29}$&$106^{+29}_{-29}$&$116^{+35}_{-31}$&$130^{+45}_{-33}$&$124^{+43}_{-30}$&$118^{+40}_{-29}$\\\hline
					$\delta R^{kl}_{\tau}(\text{Exp.})$  &$\text{}^{+23.2}_{-29.3}$&$\text{}^{+16.5}_{-19.0}$&$\text{}^{+13.7}_{-15.4}$&$\text{}^{+11.5}_{-12.8}$&$\text{}^{+9.9}_{-11.0}$&$\text{}^{+19.6}_{-24.2}$&$\text{}^{+19.1}_{-23.2}$&$\text{}^{+17.0}_{-19.9}$&$\text{}^{+13.9}_{-16.0}$&$\text{}^{+11.8}_{-13.5}$\\\hline
					$\xi\in\left[.75,2.0\right]$&$\text{}^{+19.0}_{-11.0}$&$\text{}^{+22.6}_{-12.8}$&$\text{}^{+23.7}_{-13.5}$&$\text{}^{+21.7}_{-12.3}$&$\text{}^{+19.5}_{-11.1}$&$\text{}^{+17.6}_{-10.2}$&$\text{}^{+19.8}_{-11.2}$&$\text{}^{+21.2}_{-12.0}$&$\text{}^{+19.7}_{-11.1}$&$\text{}^{+18.0}_{-10.2}$\\\hline
					Truncation uncertainty&$\text{}^{-8.0}_{+10.1}$&$\text{}^{-14.7}_{+21.7}$&$\text{}^{-21.5}_{+36.7}$&$\text{}^{-20.2}_{+34.5}$&$\text{}^{-18.9}_{+32.2}$&$\text{}^{-7.4}_{+9.3}$&$\text{}^{-12.8}_{+19.0}$&$\text{}^{-19.2}_{+33.3}$&$\text{}^{-18.5}_{+32.1}$&$\text{}^{-17.8}_{+30.7}$\\\hline
					$s_0\in\left[3,\mtsq\right](\GeV^2)$&8.7&13.2&15.7&15.3&15.4&8.0&11.5&13.9&13.8&14.3\\\hline
				\end{tabular}
				\caption{Strange quark mass using FOPT in prescription I. Other sources of uncertainties are not shown in the table but are added in the quadrature and appear for $m_s(\mtsq)$ in the second row.}
				\label{tab:msFOPT}
			\end{table}
			\begin{table}[H]
				\centering
				\begin{tabular}{|c|c|c|c|c|c|c|c|c|c|c|c|}
					\hline
					Parameter&\multicolumn{5}{c|}{Moments ALEPH\cite{Chen:2001qf}}&\multicolumn{5}{c|}{Moment OPAL\cite{OPAL:2004icu}} \\ \cline{2-6}\cline{7-11}
					\text{}&(0,0)&(1,0)&(2,0)&(3,0)&(4,0)&(0,0)&(1,0)&(2,0)&(3,0)&(4,0)\\ \hline
					$m_s(\mtsq)$&$114_{-34}^{+33}$&$107^{+30}_{-24}$&$97^{+28}_{-20}$&$87^{+27}_{-20}$&$78_{-19}^{+25}$&$106^{+29}_{-29}$&$93^{+28}_{-25}$&$85^{+26}_{-21}$&$78^{+25}_{-19}$&$72^{+24}_{-18}$\\\hline
					$\delta R^{kl}_{\tau}(\text{Exp.})$  &$\text{}^{+23.2}_{-29.3}$&$\text{}^{+13.3}_{-15.2}$&$\text{}^{+9.3}_{-10.4}$&$\text{}^{+7.6}_{-8.4}$&$\text{}^{+6.5}_{-7.1}$&$\text{}^{+19.6}_{-24.2}$&$\text{}^{+15.3}_{-18.5}$&$\text{}^{+11.4}_{-13.3}$&$\text{}^{+9.1}_{-10.4}$&$\text{}^{+7.6}_{-8.6}$\\\hline
					$\xi\in\left[.75,2.0\right]$&$\text{}^{+19.0}_{-11.0}$&$\text{}^{+20.8}_{-12.2}$&$\text{}^{+20.8}_{-12.2}$&$\text{}^{+19.8}_{-11.7}$&$\text{}^{+18.4}_{-10.8}$&$\text{}^{+17.6}_{-10.2}$&$\text{}^{+18.1}_{-10.6}$&$\text{}^{+18.2}_{-10.7}$&$\text{}^{+17.7}_{-10.4}$&$\text{}^{+16.9}_{-9.9}$\\\hline
					Truncation uncertainty&$\text{}^{-8.0}_{+10.1}$&$\text{}^{-9.0}_{+12.0}$&$\text{}^{-9.1}_{+12.7}$&$\text{}^{-8.9}_{+12.8}$&$\text{}^{-8.5}_{+12.5}$&$\text{}^{-7.4}_{+9.3}$&$\text{}^{-7.8}_{+10.5}$&$\text{}^{-8.1}_{+11.2}$&$\text{}^{-8.0}_{+11.5}$&$\text{}^{-7.8}_{+11.6}$\\\hline
					$s_0\in\left[3,\mtsq\right](\GeV^2)$&8.7&10.5&10.6&10.3&10.1&8.0&9.1&9.3&9.3&9.4\\\hline
				\end{tabular}
				\caption{ Strange quark mass using FOPT in prescription II. Other sources of uncertainties are not shown in the table but are added in the quadrature and appear for $m_s(\mtsq)$ in the second row}
				\label{tab:msFOPT2}
			\end{table}
		\end{center}
		\subsubsection{\bf{Strange quark mass determination using the RGSPT scheme}}
		The RGSPT determination of strange quark mass is presented in tables~\eqref{tab:msRGSPT},\eqref{tab:msRGSPT2} and the most crucial feature of this scheme is that it provides minimum scale uncertainty compared to the CIPT and FOPT. Another important advantage we can infer from prescription II is that it gives the lowest uncertainty among other perturbative schemes.
		\begin{center}
			\begin{table}[H]
				\centering
				\begin{tabular}{|c|c|c|c|c|c|c|c|c|c|c|c|c|}
					\hline
					Parameter&\multicolumn{5}{c|}{Moments ALEPH\cite{Chen:2001qf}}&\multicolumn{5}{c|}{Moment OPAL\cite{OPAL:2004icu}} \\ \cline{2-6}\cline{7-11}
					\text{} &(0,0)&(1,0)&(2,0)&(3,0)&(4,0)&(0,0)&(1,0)&(2,0)&(3,0)&(4,0)\\ \hline
					$m_s(\mtsq)$&$123_{-34}^{+28}$&$121^{+23}_{-23}$&$120^{+26}_{-23}$&$125^{+37}_{-27}$&$113_{-25}^{+35}$&$114^{+24}_{-28}$&$104^{+23}_{-25}$&$105^{+25}_{-24}$&$113^{+35}_{-27}$&$104^{+33}_{-25}$\\\hline
					$\delta R^{kl}_{\tau}$(\text{Exp.})&$\text{}^{+25.2}_{-32.0}$&$\text{}^{+15.2}_{-17.4}$&$\text{}^{+11.6}_{-13.0}$&$\text{}^{+10.9}_{-12.0}$&$\text{}^{+9.1}_{-10.1}$&$\text{}^{+21.3}_{-26.3}$&$\text{}^{+17.5}_{-21.2}$&$\text{}^{+14.3}_{-16.7}$&$\text{}^{+13.1}_{-15.0}$&$\text{}^{+10.8}_{-12.3}$\\\hline
					$\xi\in\left[.75,2\right]$&$\text{}^{+3.3}_{-2.1}$&$\text{}^{+4.3}_{-2.8}$&$\text{}^{+5.1}_{-3.4}$&$\text{}^{+6.2}_{-4.2}$&$\text{}^{+5.9}_{-4.0}$&$\text{}^{+3.0}_{-2.0}$&$\text{}^{+3.7}_{-2.4}$&$\text{}^{+4.6}_{-3.0}$&$\text{}^{+5.7}_{-3.8}$&$\text{}^{+5.5}_{-3.7}$\\\hline
					Truncation uncertainty&$\text{}^{-7.2}_{+8.7}$&$\text{}^{-9.7}_{+12.8}$&$\text{}^{-12.7}_{+18.5}$&$\text{}^{-18.2}_{+31.3}$&$\text{}^{-16.9}_{+29.6}$&$\text{}^{-6.7}_{+8.1}$&$\text{}^{-8.4}_{+11.1}$&$\text{}^{-11.2}_{+16.4}$&$\text{}^{-16.5}_{+28.8}$&$\text{}^{-15.8}_{+27.9}$\\\hline
					$s_0\in\left[3,\mtsq\right](\GeV^2)$&8.1&11.4&13.1&15.1&14.6&7.5&9.9&11.5&13.6&13.5\\\hline
				\end{tabular}
				\caption{ Strange quark mass using RGSPT in prescription I. Other sources of uncertainties are not shown in the table but are added in the quadrature and appear for $m_s(\mtsq)$ in the second row.}
				\label{tab:msRGSPT}
			\end{table}
			\begin{table}[H]
				\centering
				\begin{tabular}{|c|c|c|c|c|c|c|c|c|c|c|c|}
					\hline
					Parameter&\multicolumn{5}{c|}{Moments ALEPH\cite{Chen:2001qf}}&\multicolumn{5}{c|}{Moment OPAL\cite{OPAL:2004icu}} \\ \cline{2-6}\cline{7-11}
					\text{}&(0,0)&(1,0)&(2,0)&(3,0)&(4,0)&(0,0)&(1,0)&(2,0)&(3,0)&(4,0)\\ \hline
					$m_s(\mtsq)$&$123_{-34}^{+28}$&$110^{+21}_{-21}$&$95^{+18}_{-17}$&$82^{+17}_{-16}$&$70_{-14}^{+16}$&$114^{+24}_{-28}$&$95^{+21}_{-23}$&$84^{+18}_{-18}$&$74^{+17}_{-16}$&$65^{+16}_{-14}$\\\hline
					$\delta R^{kl}_{\tau}(\text{Exp.})$  &$\text{}^{+25.2}_{-32.0}$&$\text{}^{+13.9}_{-16.0}$&$\text{}^{+9.4}_{-10.4}$&$\text{}^{+7.3}_{-8.1}$&$\text{}^{+5.9}_{-6.5}$&$\text{}^{+21.3}_{-26.3}$&$\text{}^{+16.1}_{-19.4}$&$\text{}^{+11.5}_{-13.4}$&$\text{}^{+8.8}_{-10.0}$&$\text{}^{+7.0}_{-7.9}$\\\hline
					$\xi\in\left[.75,2.0\right]$&$\text{}^{+3.3}_{-2.1}$&$\text{}^{+4.0}_{-2.6}$&$\text{}^{+4.3}_{-2.8}$&$\text{}^{+4.4}_{-2.9}$&$\text{}^{+4.4}_{-2.9}$&$\text{}^{+3.0}_{-2.0}$&$\text{}^{+3.4}_{-2.3}$&$\text{}^{+3.8}_{-2.5}$&$\text{}^{+4.0}_{-2.6}$&$\text{}^{+4.1}_{-2.7}$\\\hline
					Truncation uncertainty&$\text{}^{-7.2}_{+8.7}$&$\text{}^{-8.1}_{+10.3}$&$\text{}^{-8.2}_{+10.9}$&$\text{}^{-7.9}_{+11.1}$&$\text{}^{-7.5}_{+10.9}$&$\text{}^{-6.7}_{+8.1}$&$\text{}^{-7.0}_{+8.9}$&$\text{}^{-7.2}_{+9.6}$&$\text{}^{-7.1}_{+10.0}$&$\text{}^{-6.9}_{+10.1}$\\\hline
					$s_0\in\left[3,\mtsq\right](\GeV^2)$&8.1&10.4&10.4&9.9&9.3&7.5&9.0&9.1&8.9&8.6\\\hline
				\end{tabular}
				\caption{RGSPT determination of strange quark mass using in prescription II. Only the main sources of uncertainty are shown separately, while the rest are added directly in the quadrature of $m_s(\mtsq)$.}
				\label{tab:msRGSPT2}
			\end{table}
		\end{center}

		\subsection{The strange quark mass determinations using phenomenological inputs}\label{app:pheno_mass}
		In this section, the strange quark mass is calculated using the phenomenological parametrization along with the perturbative $``L+T"-$contributions described in section~\ref{sec:rev_pheno} and section~\ref{sec:OPE_contributions}. The quark mass is calculated in the two prescriptions for various moments, and the details of sources of uncertainties are presented in the tables.
		\subsubsection{ \bf{Strange quark mass determination using CIPT scheme}}
		The CIPT determination of $m_s(\mtsq)$ in this section makes use of the full dimension-2 results of $\order{\alpha_s^3}$ as the series presented in eq.~\eqref{eq:dim2_CI} is convergent for all moments, and prescription I and prescription II yield the same determinations. The results are shown in the table~\ref{tab:msphenoCI}.

		\begin{table}[H]
			\centering
			\begin{tabular}{|c|c|c|c|c|c|c|c|c|c|c|c|c|}
				\hline
				Parameter&\multicolumn{6}{c|}{Moments ALEPH\cite{Chen:2001qf}}&\multicolumn{6}{c|}{Moment OPAL\cite{OPAL:2004icu}} \\ \cline{2-7}\cline{8-13}
				&(0,0)&(1,0)&(2,0)&(3,0)&(4,0)&$(4,0)\footnotemark[\value{footnote}]$&(0,0)&(1,1)&(2,0)&(3,0)&(4,0)&$(4,0)\footnotemark[\value{footnote}]$\\ \hline
				$m_s(\mtsq)$&$187_{-81}^{+63}$&$162_{-34}^{+31}$&$136_{-24}^{+25}$&$115_{-20}^{+25}$&$98_{-15}^{+22}$&$98_{-19}^{+25}$&$166_{-69}^{+54}$&$134_{-41}^{+34}$&$116_{-27}^{+26}$&$102_{-21}^{+24}$&$91_{-16}^{+21}$&$91_{-19}^{+24}$\\\hline
				$\delta R^{kl}_{\tau}(\text{Exp.})$  &$\text{}^{+57.2}_{-79.0}$&$\text{}^{+25.5}_{-30.1}$&$\text{}^{+15.5}_{-17.5}$&$\text{}^{+11.4}_{-12.7}$&$\text{}^{+8.8}_{-9.7}$&$\text{}^{+8.8}_{-9.7}$&$\text{}^{+49.0}_{-67.8}$&$\text{}^{+30.2}_{-39.0}$&$\text{}^{+19.4}_{-23.3}$&$\text{}^{+13.8}_{-16.0}$&$\text{}^{+10.5}_{-11.9}$&$\text{}^{+10.5}_{-11.9}$\\\hline
				$\xi\in \left[.75,2.0\right]$&$\text{}^{-9.5}_{+22.9}$&$\text{}^{-1.0}_{+9.1}$&$\text{}^{+10.8}_{-0.1}$&$\text{}^{+14.4}_{-1.7}$&$\text{}^{+16.3}_{-3.8}$&$\text{}^{+16.3}_{-3.8}$&$\text{}^{-7.9}_{+19.5}$&$\text{}_{+7.1}^{-0.7}$&$\text{}^{+9.5}_{-0.2}$&$\text{}^{+12.8}_{-1.6}$&$\text{}^{+14.8}_{-3.5}$&$\text{}^{+14.8}_{-3.5}$\\\hline
				Truncation uncertainty&$\text{}^{+3.3}_{-3.1}$&$\text{}_{+4.9}^{-4.5}$&$\text{}_{+8.4}^{-7.1}$&$\text{}_{+10.0}^{-7.9}$&$\text{}_{+10.8}^{-8.1}$&$\text{}_{-2.7}^{+2.8}$&$\text{}_{-2.7}^{+2.8}$&$\text{}^{-3.7}_{+4.0}$&$\text{}^{-6.0}_{+7.1}$&$\text{}^{-7.0}_{+8.9}$&$\text{}^{-7.5}_{+9.9}$&$\text{}^{-7.5}_{+9.9}$\\\hline
				$s_0\in\left[3,\mtsq\right](\GeV^2)$&$11.6$&$14.7$&$14.5$&$13.7$&$5.8$&$13.0$&$10.2$&$12.9$&$12.8$&$12.2$&$5.2$&$11.7$\\\hline
			\end{tabular}
			\caption{ Strange quark mass using CIPT using phenomenological inputs for the longitudinal component. Only the main sources of uncertainty are shown separately, while the rest are already added to the quadrature and appear in the total uncertainty in $m_s(\mtsq)$.}
			\label{tab:msphenoCI}
		\end{table}

		\subsubsection{\bf{Strange quark mass determination using FOPT scheme}}
		The FOPT determination of $m_s(\mtsq)$ in this section involves determination in both prescription (I-II) as the perturbation series is not well convergent for different moments, as shown in eq.~\eqref{eq:dim2_FO}. The results are presented in table~\ref{tab:msphenofixed} and table~\ref{tab:msphenofixed2}.
		\begin{center}
			\begin{table}[H]
				\centering
				\begin{tabular}{|c|c|c|c|c|c|c|c|c|c|c|}
					\hline
					Parameter&\multicolumn{5}{c|}{Moments ALEPH\cite{Chen:2001qf}}&\multicolumn{5}{c|}{Moment OPAL\cite{OPAL:2004icu}} \\ \cline{2-6}\cline{7-11} &(0,0)&(1,0)&(2,0)&(3,0)&(4,0)&(0,0)&(1,0)&(2,0)&(3,0)&(4,0)\\ \hline
					$m_s(\mtsq)$&$141_{-60}^{+46}$&$133_{-31}^{+35}$&$135_{-29}^{+38}$&$145_{-33}^{+50}$&$135_{-31}^{+46}$&$125_{-51}^{+40}$&$111_{-35}^{+34}$&$116_{-30}^{+36}$&$130_{-33}^{+46}$&$125_{-30}^{+44}$\\\hline
					$\delta R^{kl}_{\tau}\left(\mtsq\right)(\text{Exp.})$ & $\text{}^{+41.2}_{-57.9}$&$\text{}^{+20.5}_{-24.2}$&$\text{}^{+15.0}_{-16.9}$&$\text{}^{+13.8}_{-15.4}$&$\text{}^{+11.4}_{-12.7}$&$\text{}^{+35.6}_{-49.8}$&$\text{}^{+24.3}_{-31.2}$&$\text{}^{+18.7}_{-22.5}$&$\text{}^{+16.7}_{-19.5}$&$\text{}^{+13.6}_{-15.6}$\\\hline
					$\xi\in\left[.75,2\right]$&$\text{}_{-6.4}^{+16.5}$&$\text{}_{-10.5}^{+21.7}$&$\text{}^{+24.5}_{-12.0}$&$\text{}^{+25.5}_{-12.6}$&$\text{}^{+23.0}_{-11.4}$&$\text{}^{+14.8}_{-5.8}$&$\text{}^{+18.1}_{-8.8}$&$\text{}^{+21.0}_{-10.3}$&$\text{}^{+22.7}_{-11.2}$&$\text{}^{+21.1}_{-10.4}$\\\hline
					Truncation uncertainty&$\text{}^{-5.8}_{+6.7}$&$\text{}^{-9.7}_{+12.4}$&$\text{}^{-14.4}_{+21.1}$&$\text{}^{-21.2}_{+36.8}$&$\text{}^{-19.9}_{+34.4}$&$\text{}^{-5.1}_{+5.9}$&$\text{}^{-8.0}_{+10.2}$&$\text{}^{-12.4}_{+18.2}$&$\text{}^{-19.0}_{+33.4}$&$\text{}^{-18.6}_{+32.4}$\\\hline
					$s_0\in\left[3,\mtsq\right](\GeV^2)$&$10.7$&$13.1$&$14.4$&$16.2$&$16.3$&$9.5$&$11.6$&$12.8$&$14.6$&$14.9$\\\hline
				\end{tabular}
				\caption{FOPT determination of $m_s(\mtsq)$ using prescription I. Only major sources of uncertainty are shown separately, while the rest are added to the quadrature and appear in the total uncertainty in $m_s(\mtsq)$.}
				\label{tab:msphenofixed}
			\end{table}

			\begin{table}[H]
				\centering
				\begin{tabular}{|c|c|c|c|c|c|c|c|c|c|c|c|}
					\hline
					Parameter&\multicolumn{5}{c|}{Moments ALEPH\cite{Chen:2001qf}}&\multicolumn{5}{c|}{Moment OPAL\cite{OPAL:2004icu}} \\ \cline{2-6}\cline{7-11} &(0,0)&(1,0)&(2,0)&(3,0)&(4,0)&(0,0)&(1,0)&(2,0)&(3,0)&(4,0)\\ \hline
					$m_s(\mtsq)$&$141_{-60}^{+46}$&$133_{-31}^{+35}$&$121_{-33}^{+26}$&$109_{-23}^{+32}$&$99_{-22}^{+30}$&$125_{-51}^{+40}$&$111_{-35}^{+34}$&$104_{-27}^{+31}$&$97_{-23}^{+29}$&$91_{-22}^{+28}$\\\hline
					$\delta R^{kl}_{\tau}\left(\mtsq\right)(\text{Exp.})$ & $\text{}^{+41.2}_{-57.9}$&$\text{}^{+20.5}_{-24.2}$&$\text{}^{+15.0}_{-16.9}$&$\text{}^{+13.8}_{-15.4}$&$\text{}^{+11.4}_{-12.7}$&$\text{}^{+35.6}_{-49.8}$&$\text{}^{+24.3}_{-31.2}$&$\text{}^{+18.7}_{-22.5}$&$\text{}^{+16.7}_{-19.5}$&$\text{}^{+13.6}_{-15.6}$\\\hline
					$\xi\in\left[.75,2\right]$&$\text8{}_{-6.4}^{+16.5}$&$\text{}_{-10.5}^{+21.7}$&$\text{}^{+24.5}_{-12.0}$&$\text{}^{+25.5}_{-12.6}$&$\text{}^{+23.0}_{-11.4}$&$\text{}^{+14.8}_{-5.8}$&$\text{}^{+18.1}_{-8.8}$&$\text{}^{+21.0}_{-10.3}$&$\text{}^{+22.7}_{-11.2}$&$\text{}^{+21.1}_{-10.4}$\\\hline
					Truncation uncertainty&$\text{}^{-5.8}_{+6.7}$&$\text{}^{-9.7}_{+12.4}$&$\text{}^{-10.4}_{+14.5}$&$\text{}^{-10.7}_{+15.1}$&$\text{}^{-10.4}_{+15.2}$&$\text{}^{-5.1}_{+5.9}$&$\text{}^{-8.0}_{+10.2}$&$\text{}^{-9.2}_{+12.5}$&$\text{}^{-9.6}_{+13.5}$&$\text{}^{-9.7}_{+14.1}$\\\hline
					$s_0\in\left[3,\mtsq\right](\GeV^2)$&$10.7$&$13.1$&$13.3$&$13.0$&$12.8$&$9.5$&$11.6$&$11.7$&$11.6$&$11.7$\\\hline
				\end{tabular}
				\caption{ The FOPT determination of $m_s(\mtsq)$ in prescription II. Only major sources of uncertainty are shown separately, while the rest are added to the quadrature and appear in the total uncertainty in $m_s(\mtsq)$.}
				\label{tab:msphenofixed2}
			\end{table}

		\end{center}
		\subsubsection{\bf{Strange quark mass determination using RGSPT scheme}}
		The RGSPT determination in prescriptions I-II is shown in the table~\ref{tab:msphenosummed} as the $(4,0)$-moment is not term by term convergent till $\order{\alpha_s^3}$.
		\begin{center}

			\begin{table}[H]
				\centering
				\begin{tabular}{|c|c|c|c|c|c|c|c|c|c|c|c|c|c|}
					\hline
					Parameter&\multicolumn{6}{c|}{Moments ALEPH\cite{Chen:2001qf}}&\multicolumn{6}{c|}{Moment OPAL\cite{OPAL:2004icu}} \\ \cline{2-7}\cline{8-13}\hline &(0,0)&(1,0)&(2,0)&(3,0)&(4,0)&$(4,0)\footnotemark[\value{footnote}]$&(0,0)&(1,0)&(2,0)&(3,0)&(4,0)&$(4,0)\footnotemark[\value{footnote}]$\\ \hline
					$m_s(\mtsq)(\text{in}\MeV)$&$178_{-77}^{+57}$&$154_{-33}^{+29}$&$130_{-23}^{+23}$&$111_{-20}^{+21}$&$108_{-21}^{+25}$&$96_{-18}^{+20}$&$157_{-66}^{+49}$&$127_{-39}^{+32}$&$112_{-26}^{+24}$&$99_{-21}^{+21}$&$100_{-21}^{+24}$&$89_{-18}^{+19}$\\\hline
					$\delta R^{kl}_{\tau}(\text{Exp.})$  &$\text{}_{-75.5}^{+55.8}$&$\text{}_{-28.7}^{+24.5}$&$\text{}_{-16.8}^{+14.9}$&$\text{}_{-12.2}^{+11.0}$&$\text{}_{-10.6}^{+9.6}$&$\text{}_{-9.4}^{+8.6}$&$\text{}_{-64.6}^{+47.4}$&$\text{}_{-37.0}^{+28.8}$&$\text{}_{-22.3}^{+18.5}$&$\text{}_{-15.4}^{+13.3}$&$\text{}_{-12.9}^{+11.4}$&$\text{}_{-11.5}^{+10.1}$\\\hline
					$\xi\in\left[.75,2\right]$&$\text{}_{+2.6}^{-2.7}$&$\text{}_{-0.2}^{+1.1}$&$\text{}_{-1.5}^{+2.9}$&$\text{}_{-2.1}^{+3.7}$&$\text{}_{-2.7}^{+4.6}$&$\text{}_{-2.3}^{+4.0}$&$\text{}_{+2.1}^{-2.2}$&$\text{}_{-0.3}^{+1.1}$&$\text{}_{-1.4}^{+2.5}$&$\text{}_{-1.9}^{+3.3}$&$\text{}_{-2.5}^{+4.3}$&$\text{}_{-2.2}^{+3.7}$\\\hline
					Truncation uncertainty&$\text{}_{-2.7}^{+2.9}$&$\text{}_{+6.1}^{-5.4}$&$\text{}^{-7.9}_{+9.7}$&$\text{}^{-8.7}_{+11.3}$&$\text{}^{-11.6}_{+16.9}$&$\text{}^{-8.8}_{+12.1}$&$\text{}_{-2.4}^{+2.5}$&$\text{}_{+4.9}^{-4.4}$&$\text{}^{-6.8}_{+8.3}$&$\text{}^{-7.7}_{+10.1}$&$\text{}^{-10.7}_{+15.6}$&$\text{}^{-8.1}_{+11.1}$\\\hline
					$s_0\in\left[3,\mtsq\right](\GeV^2)$&$11.7$&$14.5$&$14.2$&$13.4$&$12.8$&$10.3$&$12.8$&$12.5$&$12.5$&$12.0$&$12.8$&$11.6$\\\hline
				\end{tabular}
				\caption{ Strange quark mass using RGSPT. Other sources of uncertainties are not shown in the table but are added in the quadrature and appear for $m_s(\mtsq)$ in the second row.}
				\label{tab:msphenosummed}
			\end{table}
			
		\end{center}
		
		\footnotetext[\value{footnote}]{prescription II is used for these moments.}
		
		\subsection{Details of the \texorpdfstring{$\vert V_{us}\vert$}{} determinations from OPAL data}\label{app:vus}
		The CKM matrix element $\vert V_{us}\vert$ is calculated using eq.~\eqref{eq:vus_calc} from the available data on moments for strange and non-strange components. Details of extraction from these moments and associated uncertainties from the inputs parameters are presented in this section. Purely pQCD inputs for the longitudinal component is used to extract $\vert V_{us}\vert$ in the table~\ref{tab:VusAllpert1_Opal} and table~\ref{tab:VusAll2pert_Opal}. Large theoretical uncertainties in prescription II come from the truncation of perturbative series. Determination of $\vert V_{us}\vert$ from the phenomenological inputs for longitudinal contribution is presented in the table~\ref{tab:VusAll1pheno_Opal} and table~\ref{tab:VusAll2pheno_Opal} for prescriptions I and II, respectively.\par
		\begin{table}[H]
			\centering\scalebox{.94}{
				\begin{tabular}{|c|c|c|c|c|c|c|c|c|c|c|c|c|c|c|c|}
					\hline
					Parameters&\multicolumn{5}{c|}{$\vert V_{us}\vert_{CIPT}$}&\multicolumn{5}{c|}{$\vert V_{us}\vert_{FOPT}$}&\multicolumn{5}{c|}{$\vert V_{us}\vert_{RGSPT}$}\\
					\cline{2-6}\cline{7-11}\cline{12-16}
					\text{}&(0,0)&(1,0)&(2,0)&(3,0)&(4,0)&(0,0)&(1,0)&(2,0)&(3,0)&(4,0)&(0,0)&(1,0)&(2,0)&(3,0)&(4,0)\\
					\hline
					$\vert V_{us}\vert$(central)&0.2217&0.2224&0.2219&0.2241&0.2220&0.2227&0.2208&0.2183&0.2181&0.2182&0.2221&0.2229&0.2223&0.2204&0.2217\\ \hline
					$m_s$&$\text{}^{+0.0014}_{-0.0012}$&$\text{}^{+0.0021}_{-0.0018}$&$\text{}^{+0.0025}_{-0.0022}$&$\text{}^{+0.0038}_{-0.0032}$&$\text{}^{+0.0037}_{-0.0031}$&$\text{}^{+0.0017}_{-0.0015}$&$\text{}^{+0.0017}_{-0.0014}$&$\text{}^{+0.0016}_{-0.0014}$&$\text{}^{+0.0021}_{-0.0018}$&$\text{}^{+0.0027}_{-0.0022}$&$\text{}^{+0.0015}_{-0.0013}$&$\text{}^{+0.0022}_{-0.0019}$&$\text{}^{+0.0027}_{-0.0023}$&$\text{}^{+0.0027}_{-0.0023}$&$\text{}^{+0.0037}_{-0.0031}$\\ \hline
					Experimental &$\text{}^{+0.0033}_{-0.0034}$&$\text{}^{+0.0037}_{-0.0037}$&$\text{}^{+0.0036}_{-0.0037}$&$\text{}^{+0.0037}_{-0.0038}$&$\text{}^{+0.0038}_{-0.0038}$&$\text{}^{+0.0033}_{-0.0034}$&$\text{}^{+0.0036}_{-0.0037}$&$\text{}^{+0.0036}_{-0.0036}$&$\text{}^{+0.0036}_{-0.0037}$&$\text{}^{+0.0037}_{-0.0038}$&$\text{}^{+0.0033}_{-0.0034}$&$\text{}^{+0.0037}_{-0.0037}$&$\text{}^{+0.0037}_{-0.0037}$&$\text{}^{+0.0037}_{-0.0037}$&$\text{}^{+0.0038}_{-0.0038}$\\ \hline
					Total theory&$\text{}^{+0.0017}_{-0.0017}$&$\text{}^{+0.0028}_{-0.0029}$&$\text{}^{+0.0040}_{-0.0042}$&$\text{}^{+0.0065}_{-0.0065}$&$\text{}^{+0.0078}_{-0.0074}$&$\text{}^{+0.0024}_{-0.0024}$&$\text{}^{+0.0027}_{-0.0028}$&$\text{}^{+0.0032}_{-0.0032}$&$\text{}^{+0.0042}_{-0.0041}$&$\text{}^{+0.0053}_{-0.0051}$&$\text{}^{+0.0017}_{-0.0015}$&$\text{}^{+0.0028}_{-0.0026}$&$\text{}^{+0.0039}_{-0.0036}$&$\text{}^{+0.0050}_{-0.0047}$&$\text{}^{+0.0070}_{-0.0064}$\\ \hline                    $s_0\in\left[2.50,\mtsq\right]$&0.0032&0.0070&0.0117&0.0188&0.0229&0.0042&0.0089&0.0122&0.0146&0.0200&0.0034&0.0074&0.0126&0.0204&0.0272\\\hline
					total&$\text{}^{+0.0049}_{-0.0050}$&$\text{}^{+0.0084}_{-0.0085}$&$\text{}^{+0.0129}_{-0.0129}$&$\text{}^{+0.0203}_{-0.0203}$&$\text{}^{+0.0245}_{-0.0244}$&$\text{}^{+0.0058}_{-0.0059}$&$\text{}^{+0.0100}_{-0.0100}$&$\text{}^{+0.0131}_{-0.0131}$&$\text{}^{+0.0156}_{-0.0156}$&$\text{}^{+0.0210}_{-0.0210}$&$\text{}^{+0.0051}_{-0.0050}$&$\text{}^{+0.0088}_{-0.0087}$&$\text{}^{+0.0137}_{-0.0136}$&$\text{}^{+0.0213}_{-0.0213}$&$\text{}^{+0.0283}_{-0.0282}$\\ \hline
			\end{tabular}}
			\caption{Determination of $\vert V_{us}\vert$ in various schemes from the OPAL data using prescription I. The longitudinal component is calculated using the pQCD Adler function.}
			\label{tab:VusAllpert1_Opal}
		\end{table}
		\begin{table}[H]
			\centering
			\scalebox{.94}{	\begin{tabular}{|c|c|c|c|c|c|c|c|c|c|c|c|c|c|c|c|c|c|}
					\hline
					Parameters&\multicolumn{5}{c|}{$\vert V_{us}\vert_{CIPT}$}&\multicolumn{5}{c|}{$\vert V_{us}\vert_{FOPT}$}&\multicolumn{5}{c|}{$\vert V_{us}\vert_{RGSPT}$}\\
					\cline{2-6}\cline{7-11}\cline{12-16}
					&(0,0)&(1,0)&(2,0)&(3,0)&(4,0)&(0,0)&(1,0)&(2,0)&(3,0)&(4,0)&(0,0)&(1,0)&(2,0)&(3,0)&(4,0)\\
					\hline
					$\vert V_{us}\vert$(central)&0.2217&0.2239&0.2275&0.2341&0.2452&0.2227&0.2246&0.2271&0.2310&0.2365&0.2221&0.2246&0.2286&0.2358&0.2477\\ \hline
					$m_s$&$\text{}^{+0.0014}_{-0.0012}$&$\text{}^{+0.0025}_{-0.0021}$&$\text{}^{+0.0043}_{-0.0036}$&$\text{}^{+0.0072}_{-0.0059}$&$\text{}^{+0.0126}_{-0.0097}$&$\text{}^{+0.0017}_{-0.0015}$&$\text{}^{+0.0028}_{-0.0024}$&$\text{}^{+0.0042}_{-0.0035}$&$\text{}^{+0.0062}_{-0.0051}$&$\text{}^{+0.0090}_{-0.0072}$&$\text{}^{+0.0015}_{-0.0013}$&$\text{}^{+0.0027}_{-0.0023}$&$\text{}^{+0.0047}_{-0.0039}$&$\text{}^{+0.0079}_{-0.0064}$&$\text{}^{+0.0138}_{-0.0105}$\\ \hline
					Experimental &$\text{}^{+0.0033}_{-0.0034}$&$\text{}^{+0.0037}_{-0.0038}$&$\text{}^{+0.0037}_{-0.0038}$&$\text{}^{+0.0039}_{-0.0040}$&$\text{}^{+0.0042}_{-0.0043}$&$\text{}^{+0.0033}_{-0.0034}$&$\text{}^{+0.0037}_{-0.0038}$&$\text{}^{+0.0037}_{-0.0038}$&$\text{}^{+0.0039}_{-0.0039}$&$\text{}^{+0.0041}_{-0.0041}$&$\text{}^{+0.0033}_{-0.0034}$&$\text{}^{+0.0037}_{-0.0038}$&$\text{}^{+0.0038}_{-0.0038}$&$\text{}^{+0.0039}_{-0.0040}$&$\text{}^{+0.0043}_{-0.0043}$\\ \hline
					Total theory&$\text{}^{+0.0017}_{-0.0017}$&$\text{}^{+0.0033}_{-0.0035}$&$\text{}^{+0.0063}_{-0.0065}$&$\text{}^{+0.0121}_{-0.0116}$&$\text{}^{+0.0237}_{-0.0206}$&$\text{}^{+0.0024}_{-0.0024}$&$\text{}^{+0.0043}_{-0.0043}$&$\text{}^{+0.0071}_{-0.0068}$&$\text{}^{+0.0110}_{-0.0102}$&$\text{}^{+0.0165}_{-0.0147}$&$\text{}^{+0.0017}_{-0.0015}$&$\text{}^{+0.0033}_{-0.0030}$&$\text{}^{+0.0061}_{-0.0055}$&$\text{}^{+0.0109}_{-0.0096}$&$\text{}^{+0.0200}_{-0.0168}$\\ \hline
					$s_0\in\left[2.50,\mtsq\right]$&0.0030&0.0070&0.0117&0.0188&0.0313&0.0042&0.0089&0.0143&0.0216&0.0325&0.0032&0.0074&0.0126&0.0204&0.0338\\\hline
					total&$\text{}^{+0.0048}_{-0.0048}$&$\text{}^{+0.0086}_{-0.0087}$&$\text{}^{+0.0138}_{-0.0139}$&$\text{}^{+0.0227}_{-0.0225}$&$\text{}^{+0.0395}_{-0.0377}$&$\text{}^{+0.0058}_{-0.0059}$&$\text{}^{+0.0106}_{-0.0162}$&$\text{}^{+0.0164}_{-0.0163}$&$\text{}^{+0.0245}_{-0.0242}$&$\text{}^{+0.0368}_{-0.0359}$&$\text{}^{+0.0049}_{-0.0049}$&$\text{}^{+0.0089}_{-0.0089}$&$\text{}^{+0.0145}_{-0.0145}$&$\text{}^{+0.0235}_{-0.0229}$&$\text{}^{+0.0395}_{-0.0380}$\\ \hline
			\end{tabular}}
			\caption{Determination of $\vert V_{us}\vert$ in different schemes from the OPAL data using prescription II. The longitudinal component is calculated using the pQCD Adler function.}
			\label{tab:VusAll2pert_Opal}
		\end{table}
		\begin{table}[H]
			\centering
			\scalebox{.94}{\begin{tabular}{|c|c|c|c|c|c|c|c|c|c|c|c|c|c|c|c|}
					\hline
					Parameters&\multicolumn{5}{c|}{$\vert V_{us}\vert_{CIPT}$}&\multicolumn{5}{c|}{$\vert V_{us}\vert_{FOPT}$}&\multicolumn{5}{c|}{$\vert V_{us}\vert_{RGSPT}$}\\
					\cline{2-6}\cline{7-11}\cline{12-16}
					\text{}&(0,0)&(1,0)&(2,0)&(3,0)&(4,0)&(0,0)&(1,0)&(2,0)&(3,0)&(4,0)&(0,0)&(1,0)&(2,0)&(3,0)&(4,0)\\
					\hline
					$\vert V_{us}\vert$(central)&0.2211&0.2210&0.2212&0.2228&0.2226&0.2222&0.2224&0.2225&0.2208&0.2177&0.2213&0.2213&0.2219&0.2238&0.2233\\ \hline
					$m_s$&$\text{}^{+0.0005}_{-0.0004}$&$\text{}^{+0.0010}_{-0.0009}$&$\text{}^{+0.0018}_{-0.0015}$&$\text{}^{+0.0029}_{-0.0025}$&$\text{}^{+0.0036}_{-0.0030}$&$\text{}^{+0.0008}_{-0.0007}$&$\text{}^{+0.0014}_{-0.0012}$&$\text{}^{+0.0022}_{-0.0019}$&$\text{}^{+0.0024}_{-0.0021}$&$\text{}^{+0.0022}_{-0.0019}$&$\text{}^{+0.0005}_{-0.0004}$&$\text{}^{+0.0011}_{-0.0009}$&$\text{}^{+0.0019}_{-0.0017}$&$\text{}^{+0.0032}_{-0.0028}$&$\text{}^{+0.0038}_{-0.0032}$\\ \hline
					Experimental &$\text{}^{+0.0033}_{-0.0034}$&$\text{}^{+0.0036}_{-0.0037}$&$\text{}^{+0.0036}_{-0.0037}$&$\text{}^{+0.0037}_{-0.0038}$&$\text{}^{+0.0038}_{-0.0038}$&$\text{}^{+0.0033}_{-0.0034}$&$\text{}^{+0.0037}_{-0.0037}$&$\text{}^{+0.0037}_{-0.0037}$&$\text{}^{+0.0037}_{-0.0037}$&$\text{}^{+0.0037}_{-0.0038}$&$\text{}^{+0.0033}_{-0.0034}$&$\text{}^{+0.0036}_{-0.0037}$&$\text{}^{+0.0036}_{-0.0037}$&$\text{}^{+0.0037}_{-0.0038}$&$\text{}^{+0.0038}_{-0.0039}$\\ \hline
					Total theory&$\text{}^{+0.0005}_{-0.0005}$&$\text{}^{+0.0010}_{-0.0010}$&$\text{}^{+0.0020}_{-0.0020}$&$\text{}^{+0.0036}_{-0.0037}$&$\text{}^{+0.0053}_{-0.0054}$&$\text{}^{+0.0009}_{-0.0009}$&$\text{}^{+0.0019}_{-0.0019}$&$\text{}^{+0.0033}_{-0.0033}$&$\text{}^{+0.0039}_{-0.0039}$&$\text{}^{+0.0043}_{-0.0041}$&$\text{}^{+0.0005}_{-0.0004}$&$\text{}^{+0.0011}_{-0.0010}$&$\text{}^{+0.0022}_{-0.0020}$&$\text{}^{+0.0040}_{-0.0036}$&$\text{}^{+0.0055}_{-0.0050}$\\ \hline
					$s_0\in\left[2.5,\mtsq\right]$&0.0032 & 0.0070&0.0117&0.0188&0.0229&0.0042&0.0089&0.0122&0.0146&0.0200&0.0034&0.0074&0.0126&0.0204&0.0272\\
					\hline
					total&$\text{}^{+0.0047}_{-0.0047}$&$\text{}^{+0.0080}_{-0.0080}$&$\text{}^{+0.0124}_{-0.0124}$&$\text{}^{+0.0195}_{-0.0195}$&$\text{}^{+0.0238}_{-0.0239}$&$\text{}^{+0.0054}_{-0.0054}$&$\text{}^{+0.0098}_{-0.0099}$&$\text{}^{+0.0131}_{-0.0132}$&$\text{}^{+0.0155}_{-0.0155}$&$\text{}^{+0.0208}_{-0.0207}$&$\text{}^{+0.0048}_{-0.0048}$&$\text{}^{+0.0084}_{-0.0084}$&$\text{}^{+0.0133}_{-0.0133}$&$\text{}^{+0.0211}_{-0.0211}$&$\text{}^{+0.0280}_{-0.0279}$\\ \hline
			\end{tabular}}
			\caption{Determination of $\vert V_{us}\vert$ in different schemes from the OPAL data using prescription I. The phenomenological contributions are used for longitudinal components.}
			\label{tab:VusAll1pheno_Opal}
		\end{table}
		\begin{table}[H]
			\centering
			\scalebox{.94}{\begin{tabular}{|c|c|c|c|c|c|c|c|c|c|c|c|c|c|c|c|}
					\hline
					Parameters&\multicolumn{5}{c|}{$\vert V_{us}\vert_{CIPT}$}&\multicolumn{5}{c|}{$\vert V_{us}\vert_{FOPT}$}&\multicolumn{5}{c|}{$\vert V_{us}\vert_{RGSPT}$}\\
					\cline{2-6}\cline{7-11}\cline{12-16}
					\text{ }&(0,0)&(1,0)&(2,0)&(3,0)&(4,0)&(0,0)&(1,0)&(2,0)&(3,0)&(4,0)&(0,0)&(1,0)&(2,0)&(3,0)&(4,0)\\
					\hline
					$\vert V_{us}\vert$(central)&0.2211&0.2210&0.2212&0.2228&0.2260&0.2222&0.2224&0.2225&0.2236&0.2253&0.2213&0.2213&0.2219&0.2238&0.2274\\ \hline
					$m_s$&$\text{}^{+0.0005}_{-0.0004}$&$\text{}^{+0.0010}_{-0.0009}$&$\text{}^{+0.0018}_{-0.0015}$&$\text{}^{+0.0029}_{-0.0025}$&$\text{}^{+0.0039}_{-0.0046}$&$\text{}^{+0.0008}_{-0.0007}$&$\text{}^{+0.0014}_{-0.0012}$&$\text{}^{+0.0022}_{-0.0019}$&$\text{}^{+0.0032}_{-0.0027}$&$\text{}^{+0.0045}_{-0.0038}$&$\text{}^{+0.0005}_{-0.0004}$&$\text{}^{+0.0011}_{-0.0009}$&$\text{}^{+0.0019}_{-0.0017}$&$\text{}^{+0.0032}_{-0.0028}$&$\text{}^{+0.0051}_{-0.0043}$\\ \hline
					Experimental &$\text{}^{+0.0033}_{-0.0034}$&$\text{}^{+0.0036}_{-0.0037}$&$\text{}^{+0.0036}_{-0.0037}$&$\text{}^{+0.0037}_{-0.0038}$&$\text{}^{+0.0038}_{-0.0039}$ &$\text{}^{+0.0033}_{-0.0034}$&$\text{}^{+0.0037}_{-0.0037}$&$\text{}^{+0.0037}_{-0.0037}$&$\text{}^{+0.0037}_{-0.0038}$&$\text{}^{+0.0038}_{-0.0039}$&$\text{}^{+0.0033}_{-0.0034}$&$\text{}^{+0.0036}_{-0.0037}$&$\text{}^{+0.0036}_{-0.0037}$&$\text{}^{+0.0037}_{-0.0038}$&$\text{}^{+0.0039}_{-0.0039}$\\ \hline
					Total theory&$\text{}^{+0.0005}_{-0.0005}$&$\text{}^{+0.0010}_{-0.0010}$&$\text{}^{+0.0020}_{-0.0020}$&$\text{}^{+0.0036}_{-0.0037}$&$\text{}^{+0.0062}_{-0.0065}$&$\text{}^{+0.0009}_{-0.0009}$&$\text{}^{+0.0019}_{-0.0019}$&$\text{}^{+0.0033}_{-0.0033}$&$\text{}^{+0.0051}_{-0.0050}$&$\text{}^{+0.0074}_{-0.0071}$&$\text{}^{+0.0005}_{-0.0004}$&$\text{}^{+0.0011}_{-0.0010}$&$\text{}^{+0.0022}_{-0.0020}$&$\text{}^{+0.0040}_{-0.0036}$&$\text{}^{+0.0068}_{-0.0061}$\\ \hline
					$s_0\in\left[2.5,\mtsq\right]$&0.0030 &0 .0070&0.0117&0.0188&0.0313&0.0042&0.0089&0.0143&0.0216&0.0325&0.0032&0.0074&0.0126&0.0204&0.0338\\
					\hline
					total&$\text{}^{+0.0045}_{-0.0045}$&$\text{}^{+0.0080}_{-0.0080}$&$\text{}^{+0.0124}_{-0.0124}$&$\text{}^{+0.0195}_{-0.0196}$&$\text{}^{+0.0321}_{-0.0322}$&$\text{}^{+0.0054}_{-0.0054}$&$\text{}^{+0.0098}_{-0.0099}$&$\text{}^{+0.0151}_{-0.0151}$&$\text{}^{+0.0225}_{-0.0225}$&$\text{}^{+0.0336}_{-0.0335}$&$\text{}^{+0.0046}_{-0.0047}$&$\text{}^{+0.0084}_{-0.0084}$&$\text{}^{+0.0133}_{-0.0133}$&$\text{}^{+0.0211}_{-0.0211}$&$\text{}^{+0.0347}_{-0.0346}$\\ \hline
			\end{tabular}}
			\caption{Determination of $\vert V_{us}\vert$ from OPAL data using prescription II. The phenomenological contributions are used for longitudinal components.}
			\label{tab:VusAll2pheno_Opal}
		\end{table}

	\end{widetext}

\end{document}